\documentstyle[12pt]{article}

\begin{document}

\hoffset = -1truecm
\voffset = -2truecm
\baselineskip = 10 mm

\title{\bf Contributions of gluon recombination to saturation phenomena }

\author{
{\bf Wei Zhu}, {\bf Jianhong Ruan}, {\bf Jifeng Yang} and 
{\bf Zhenqi Shen} \\
\normalsize Department of Physics, East China Normal University,
Shanghai 200062, P.R. China \\
}

\date{}

\newpage

\maketitle

\vskip 3truecm

\begin{abstract}

	Parton distributions in the small $x$ region are numerically
predicted by using a modified DGLAP equation with the GRV-like input distributions.
We find that gluon recombination at twist-4 level obviously 
suppresses the rapid growth of parton densities with $x$ decrease.
We show that before the saturation scale $Q^2_s$ is reached, saturation and partial saturation appear 
in the small $x$ behavior of parton distributions in nucleus and free proton, respectively.
The antishadowing contributions to the saturation phenomena are also discussed.

\end{abstract}


\newpage
\begin{center}
\section{Introduction}
\end{center}

    The QCD evolution equations for parton densities at twist-2 level, both
the DGLAP equation [1] and BFKL equation [2] predict a rapid increase of the
parton densities in the small $x$ region due to parton splitting,
and the unitarity limit is violated. Therefore, the corrections
of the higher order QCD effects, which suppress or shadow the growth of parton
densities, become a focus of intensive study in recent
years.  An important character at the small $x$ limit is that the suppressed parton distributions approach
a limiting form and unitarity is restored. This is called saturation.

    There are various ways to define and analyze the saturation phenomena
based on perturbative QCD [3,4]. The shadowing corrections of gluon recombination to the integrated parton
distributions were mainly studied by adding nonlinear terms in the DGLAP evolution equation in the
collinear factorization scheme. A pioneering work in this aspect was
derived by Gribov, Levin and Ryskin in [5] and by Mueller and Qiu [6] at the twist-4 level.
The GLR-MQ equation sums the contributions of gluon recombination diagrams using the AGK
(Abramovsky, Gribov, Kancheli) cutting rule [7]. In the next step, the contributions of multi-parton correlations
are summed by using the Glauber model in Mueller's
works [8] and this Glauber-Mueller equation reduces to the GLR-MQ equation at the twist-4 approximation.

    Recently, the predictions of the GLR-MQ equation for the gluon saturation scale were studied
in [9]. However, the applications of the AGK cutting rule in the GLR-MQ equation was argued in a more general
consideration by two of us (WZ and JHR) in [10], where
the Feynman diagrams are summed in a quantum field theory framework instead of the AGK cutting rule. We shall
refer to this evolution equation as the MD-DGLAP equation. A major difference among the above mentioned
nonlinear equations is that the momentum conservation
is restored in the MD-DGLAP equation by the antishadowing corrections, which may change
the predictions of the GLR-MQ equations.

    The purpose of this work is to study the behavior of the parton (quark and gluon) distributions in $Q^2$ and $x$
at high gluon density using the MD-DGLAP equation.
It is know that the solutions of QCD evolution equations are sensitive to the input
parton distributions. Works [9] use the CTEQ input distributions [11] at $Q^2_0=1.4 GeV^2$
to fit the HERA data for the structure function $F_2(x,Q^2)$ of proton. Then
they predict the saturation scales $Q^2_s$ in proton and nucleus by evolving backward
from a higher $Q^2$ scale, where the nonlinear terms in the equation can be neglected. Therefore, one can
linearly add the input distributions of the nucleons in a nuclear target.
However, the backward evolution paths for gluon and sea quarks are not unique and hence there are uncertainties in
the results. Different from works [9], we use the GRV model [12] for input distributions, where the evolution
begins in a very low $Q^2_0< 1~GeV^2$. All partons in the GRV input take the valence-like form and
it implies the finiteness of parton number and low density of partons. Therefore, we can construct
the input distributions in the nucleus using the input set for proton and
evolve them according to the standard evolution technique.

    We fit the parameters in the input distributions using the HERA data [13] for both $F_2(x,Q^2)$ and
$dF_2(x,Q^2)/d\ln Q^2$ in proton in a limiting kinematical region.
Then we predict the small $x$ behaviors of parton
distributions at different scale $Q^2$ in proton and nucleus, respectively.  We find obvious screening effects in
quark and gluon distributions. We also show that a partial saturation appears in the parton distributions of
proton. In particular, a flatter plateau appears
in the region of smaller $x$ and lower $Q^2$ in middle and heavy nuclei. However, we have not found the 
saturation scale $Q^2_s$ in the expected domain
according to the definition in literatures [5,14]. We introduce the new scales $Q^2_R$ and $x_s$ to
describe the small $x$ behaviors of parton distributions in the leading recombination region.
We also study the contributions of the antishadowing terms in the MD-DGLAP equation.
Antishadowing compensates the lost momentum in shadowing. Although this lost momentum is little,
our calculations show that the contributions of the antishadowing terms to the saturation phenomena can not be neglected.

       The paper is organized as follows. In section 2 we introduce
the MD-DGLAP equation and compare it with the GLR-MQ equation. In section 3
we fit the parameters in the GRV-like input distributions in the proton
using the MD-DGLAP equation and HERA data, then we predict the parton distributions beyond the HERA region.
In section 4 we discuss the parton distributions in the nucleus.
The discussions and a summary are given in section 5.

\newpage
\begin{center}
\section{The MD-DGLAP equation}
\end{center}

        As we know that the DGLAP equation produces a rapid
growth of gluon density toward smaller values of $x$. The gluons therefore must begin to
spatially overlap and recombine in a thin target disc.
In works [10],
the corrections of parton
recombination to the QCD evolution equation are considered by summing up all
possible twist-4 cut diagrams in the LLA($Q^2$). In the derivation of the equation, the
time-ordered perturbation theory instead of the AGK cutting rule is used to pick up the
contributions of the leading recombination diagrams.
In consequence, the corrections of the
gluon recombination to the evolution of parton distributions with $Q^2$
are described by the following modified DGLAP equation [10]

$$\frac{dxG(x,Q^2)}{d\ln Q^2}$$
$$=P^{AP}_{gg}\otimes G(x,Q^2) + P^{AP}_{gq}\otimes S(x,Q^2)  $$
$$+\frac{\alpha_s^2K}{Q^2}
\int_{x/2}^xdx_1xx_1G^2(x_1, Q^2)
\sum_iP_i^{gg\rightarrow g}(x_1,x) $$
$$-\frac{\alpha_s^2K}{Q^2}
\int_{x}^{1/2}dx_1xx_1G^2(x_1, Q^2)
\sum_iP_i^{gg\rightarrow g}(x_1,x)
\eqno(1a)$$
for gluon distribution and

$$\frac{dxS(x,Q^2)}{d\ln Q^2}$$
$$=P^{AP}_{qg}\otimes G(x,Q^2) + P^{AP}_{qq}\otimes S(x,Q^2) $$
$$+\frac{\alpha_s^2K}{Q^2}
\int_{x/2}^xdx_1xx_1G^2(x_1, Q^2)
\sum_iP_i^{gg\rightarrow q}(x_1,x)  $$
$$-\frac{\alpha_s^2K}{Q^2}
\int_{x}^{1/2}dx_1xx_1G^2(x_1, Q^2)
\sum_iP_i^{gg\rightarrow q}(x_1,x)
\eqno(1b)$$
for sea quark distributions, where $P^{AP}$ are the evolution kernels of
the linear DGLAP equation and the recombination functions

$$\sum_iP_i^{gg\rightarrow g}(x_1,x)$$
$$=\frac{27}{64}
\frac{(2x_1-x)(-136xx_1^3-64x_1x^3+132x_1^2x^2+99x_1^4+16x^4)}
{xx_1^5}, \eqno(1c)$$
and

$$\sum_iP_i^{gg\rightarrow q}(x_1,x)$$
$$=\frac{1}{48}
\frac{(2x_1-x)(36x_1^3+49x_1x^2-14x^3-60x^2x)}{x_1^5}.
 \eqno(1d)$$

The nonlinear coefficient $K$ in Eq. (1) depends on the definition
of double parton distribution and the geometric distributions of partons inside
target. For simplicity, we take $K$ as a free parameter in this work. Comparing with the GLR-MQ equation [6]:

$$\frac{dxS(x,Q^2)}{d\ln Q^2}$$
$$=P^{AP}_{qg}\otimes G(x,Q^2) + P^{AP}_{qq}\otimes S(x,Q^2) $$
$$-\frac{1}{30}\frac{\alpha_s^2K_{GLR-MQ}}{Q^2}[xG(x,Q^2)]^2+...+G_{HT}, \eqno(2a)$$
and

$$\frac{dxG(x,Q^2)}{d\ln Q^2}
=P^{AP}_{gg}\otimes G(x,Q^2) + P^{AP}_{gq}\otimes S(x,Q^2)$$
$$-\frac{\alpha_s^2K_{GLR-MQ}}{Q^2}
\int_{x}^{1/2}\frac{dx_1}{x_1}[x_1G(x_1, Q^2)]^2, \eqno(2b)$$
there are following properties in Eq. (1):

(i) The third term on the right-hand side
of Eq. (1) is positive and it is called as the antishadowing,
while the negative fourth term arises
from the shadowing correction. The coexistence of shadowing and antishadowing in
the QCD evolution of the parton densities is a general requisition of the
local momentum conservation.
We emphasize that the shadowing and antishadowing terms are defined on
different kinematics domains $[x,1/2]$ and $[x/2, x]$, respectively.
Thus, the net recombination effects in Eq. (1) are not only related
to the value of gluon density, but also depend on the slope of the gluon distribution
in the space $[x/2, x]$, i.e., a steeper (or flatter) gluon distribution has a stronger
(or weaker) antishadowing effect.
On the other hand, the AGK cutting rule is used in the derivation of Eq. (2) and where it is assumed that both
the positive
and negative contributions of the Feynman diagrams have the same kinematics domain. This results in the
violation of the momentum conservation in the GLR-MQ equation.

(ii) The GLR-MQ equation (2) takes the double leading logarithmic
approximation (DLLA) for both $Q^2$ and $1/z$, where one keeps only
the $\ln(Q^2/\mu^2)\ln(1/z)$ factor in the solutions of the evolution
equation or, equivalently, takes only the terms having $1/z=x_1/x$ factor.
On the other hand, the MD-DGLAP equation (1) is derived under the leading logarithmic
approximation for $Q^2$. Therefore, Eqs. (1c) and (1d) contain
the terms beyond the leading $1/z$ approximation. One can find that the contributions
from these terms do not vanish even in the small $x$ region, since $z$ runs from
$x$ to $1$ as $x_1$ varies, that is $z$ is not restricted in a smaller-z region.

(iii) Note that the sea quark evolution in Eqs. (1) and (2) take different forms.
The reason is that the transition of gluon$\rightarrow$ quarks is suppressed
in the DLLA-manner.
The DLLA diagram only contains the gluon ladders and any transitions of
gluon $\rightarrow$ quark break the gluon ladder-structure.
Therefore, a special technique is used to include the
corrections of gluon recombination to the quark distributions
in the GLR-MQ equation. However, this extra technique is unnecessary in the derivation of the
MD-DGLAP equation, since we can produce the evolution equations for gluon and sea quarks
in a same framework at the $LLA(Q^2)$.

    Therefore, we use the MD-DGLAP equation to study the saturation phenomena instead of the GLR-MQ
dynamics. We shall show the different predictions of the evolution equations with or without
antishadowing terms in Sec. 5.

\newpage
\begin{center}
\section{The solutions of the MD-DGLAP equation in proton}
\end{center}

    The numerical solutions of evolution equation depend sensitively on the input parton
(gluon and quark) distributions at a lower scale $Q^2_0$. In principle, they are not calculable
within perturbative QCD but are determined by data. Because the electromagnetic probe can not directly measure
the gluon density, the input gluon distribution has a larger uncertainty. The early data for
the DIS structure functions can always be fitted by using the linear DGLAP equation provided
that a satisfying input gluon density is assumed. However, new HERA data in small $x$ region, in particular the slopes
of the structure function $dF_2/d\ln Q^2$ have restricted the above mentioned
uncertainty. In fact, the global analysis of the HERA data using the DGLAP dynamics are given
by the MRST [15] and CTEQ [11] collaborations.
Where a MRST2001 set shows the negative value of gluon distribution in $Q^2<1 GeV^2$.
It means that if a positive input gluon distribution below 1 $GeV^2$ is taken, the screening
corrections to the evolution equation are useful.

    In this work, we use the GRV-like input distributions. In the GRV model [12], the parton distributions are evolved
from a very lower resolution scale (but larger than
the QCD parameter $\Lambda_{QCD}$). A specific assumption of the GRV model is that the input
parton distributions take the simple valence-like distribution form. The GRV model with the linear DGLAP evolution
equation gives a good description for proton structure function $F_2(x,Q^2)$ in a broad region but the
fit in the slope of $F_2(x,Q^2)$ at lower $Q^2$
remains to be improved
(see the dashed curves in Fig. 2). We shall show that the corrections of the
gluon recombination improved the fit.
In practice, we use the
initial valence quark- and gluon-densities in the GRV98LO set [16]
as the input distributions
at $Q^2_0=0.34 GeV^2$, i.e.,

$$xU(x,Q^2_0)=1.239x^{0.48}(1-x)^{2.72}(1-1.8\sqrt{x}+9.5x), $$
$$xD(x,Q^2_0)=0.614(1-x)^{0.9}xU(x,Q^2_0),  $$
and

$$xG(x,Q^2_0)=17.47x^{1.6}(1-x)^{3.8}, \eqno(3)$$
and the mass of charm quark is 1.4 GeV. In the meantime, we let the parameters in
the sea quarks distributions to be determined by the HERA data [13] in the MD-DGLAP evolution
equation. The results are

$$x(\bar{d}(x,Q^2_0)+\bar{u}(x,Q^2_0))=0.9x^{0.01}(1-x)^{8.0}(1-3.6\sqrt{x}+7.8x), $$
$$x(\bar{d}(x,Q^2_0)-\bar{u}(x,Q^2_0))=0.23x^{0.48}(1-x)^{11.3}(1-12.0\sqrt{x}+50.9x). \eqno(4)$$
In this fit, the nonlinear coefficient $K$ in the evolution
equation is taken as $K=0.0014$. It implies that the nonlinear recombination corrections can not
be neglected in the HERA data [13].  The results are shown in Figs. 1 and 2 (solid curves), where
the dashed lines are the fitting results of the linear DGLAP equation using the GRV98LO [16] as
the input distributions.
One can find that the contributions of gluon recombination improve
the fit in $Q^2> 1GeV^2$. There are derivations between the fit and data in $Q^2< 1GeV^2$,
they imply that the corrections of beyond LLA($Q^2$) are necessary at very lower $Q^2$. However, as 
we shall show that our main conclusions in this work are insensitive to the choices of parameters in the 
GRV-like input.

    Now we give the predictions of the gluon and quark distributions beyond HERA region
in Figs. 3-6 (solid curves).  The parton distributions from
the DGLAP equation with GRV98LO are also plotted (dashed curves) for comparison. The results show
obvious suppression in both the quark and gluon distributions and a flat tendency
at the small $x$ limit.  Figures 5 and 6 give
the $Q^2$-dependence of parton distributions at fixed values of $x$.
The dashed curves (the predictions of the
DGLAP dynamics) in Figs. 3 and 4 have the exponential form ($\sim \exp[\sqrt{\ln 1/x}]$) and it violates the unitarity.
However, the solid curves (the results of the MD-DGLAP equation) show that the growth of $xG(x,Q^2)$ and $xS(x,Q^2)$
is slower than $\ln(1/x)$ in $1<Q^2<10 GeV^2$ and $x<10^{-6}$.
This partial saturation behavior is softer
than the predicted result by the DGLAP equation and the BFKL
equation ($\sim x^{-\lambda}$, $\lambda>0$).

    In hadron-hadron cross section, such as p-p, the Froissart boundary [17] requires

$$\sigma^{pp}(s)\le \frac{\pi}{m_{\pi}^2}log^2(s/s_0), \eqno(5)$$
at the high $s$ limit, where $s$ is the center-of-mass energy squared and $x=Q^2/s$.
The high energy limit implies the small $x$ approximation when $Q^2$ is fixed to be a few $GeV^2$.
Therefore, although the gluon- and quark-distributions in proton do not saturate at small values of $x$, but
their partial saturation-behavior satisfies the Froissart boundary
in the perturbative QCD means.

    As we know that the saturation scale $Q^2_s(x)$, which indicates the saturation limit, is
usually defined [5, 14] to be

$$\frac{dxG(x,Q^2)}{d\ln Q^2}\left \vert_{Q^2_s}\right .=0, \eqno(6a)$$

and

$$\frac{dxS(x,Q^2)}{d\ln Q^2}\left \vert_{Q^2_s}\right .=0, \eqno(6b)$$
or equivalently

$$W_s\equiv \frac{nonlinear\; terms\; of\; Eq. (1)}{linear\; terms\;of\; Eq. (1)}\left \vert_{Q^2_s}\right .= 1, \eqno(7)$$
which requires that the nonlinear recombination effect in the MD-DGLAP equation fully balances the
linear splitting effect. Thus the saturation limit is reached. However, we have not found such saturation solution
$Q^2_s$ in Eq. (1).
It means that the net shadowing (i.e., shadowing-antishadowing) effect in the leading recombination approximation
is not large enough to cancel the increase of parton densities with increasing $Q^2$. In other words, the higher-order
recombination contributions
would become significant and should be included in the evolution equation near the saturation limit.
However, in this paper we focus on the range where the gluon recombination begins to work. Therefore,
we introduce the recombination scale $Q^2_R(x)$ instead of $Q^2_s(x)$ as follows:

$$W_R\equiv \frac{nonlinear\;terms\;of\;Eq. (1)}{linear\;terms\;of\;Eq. (1)}\left \vert_{Q^2_R}\right .=
 \alpha_s(Q^2_R(x)). \eqno(8)$$
The $Q^2_R(x)$ for the gluon
distribution in proton is shown in Fig. 7 (solid curve).
It is interesting that the line has a corner near $Q^2_R(x)\simeq 2 GeV^2$. The anomalous behavior of the line at 
$Q^2_R(x)\leq 2 GeV^2$ is the results of the antishadowing corrections. In fact, the relative stronger antishadowing
effect at lower $Q^2$ locally raises the gluon distribution. It shifts the flatter plateau toward a larger $x$-
value. We can also see this effect in the behaviors of $x_s$ in Figs. 3, 8 and 9 (see following sections).

We can find that the gluon recombination obviously suppresses the evolution of parton densities in
$x<10^{-6}$ in proton.
According to the definition, one can see $Q^2_R(x)<Q^2_s(x)$. The evolution of the parton distributions from
$Q^2_R(x)$ to $Q^2_s(x)$ is a complicated process. We shall show that the contributions of the leading
recombination are also important in the nuclear target near $Q^2_R$.

\newpage
\begin{center}
\section{The solutions of the MD-DGLAP equation in nucleus}
\end{center}

 	    Nuclear target is an ideal laboratory for the research of the saturation phenomena, since the gluon recombination
corrections are enhanced due to the correlation of gluons belonging to different nucleons at
a same impact in the nuclear target.
It is well known that the nuclear shadowing is a complicated phenomenon
which has two different sources [18]: (i) it perturbatively originates from gluon recombination in the infinite
momentum frame of nuclear target, or from
multiple scattering in the target-rest frame; (ii) the non-perturbative nuclear effects.
The former is expressed as the nonlinear QCD evolution equations, while the later
relates to the structure of the input parton distributions.  
We have interest in the separate relations of nuclear saturation phenomena with the above mentioned two sources.
Therefore, at first step, we neglect the non-perturbative nuclear effects.
The gluon distribution (3)
implies a small total number and very low density of gluons, where the recombination corrections are negligible.
This conclusion can be
confirmed from the following Fig. 8, which indicates that the gluon distributions with and without recombination
corrections are similar near the evolution start point. Thus, we can
predict the nuclear parton densities using the input distributions (3-4) and the MD-DGLAP equation, where the nonlinear
terms are multiplied by $A^{1/3}$.

    Figures 8-11 and 12-15 are the results corresponding to Figs. 3-6 but for Ca($A=40$)
and Pb($A=208$), respectively.
Our results show the plateau in the parton distributions in $x<10^{-6}$ and $1<Q^2<10 GeV^2$. We note that
in the DGLAP dynamics the generated gluon
and sea quark distributions in small $x$ region will speed up transfer from the valence form
into the power form also through a plateau. However, this plateau
exists in a very narrow $Q^2$-window and it is unstable. The plateau in Figs. 8-9 and 12-13 implies
a saturation behavior in the parton distributions at the small $x$ limit. We use $x_s$
to indicate this saturation effect and is defined as

$$\frac{dxG(x,Q^2)}{d\ln \frac{1}{x}}\left \vert_{x_s}\right .=0, \eqno(9a)$$
and

$$\frac{dxq(x,Q^2)}{d\ln \frac{1}{x}}\left \vert_{x_s}\right .=0. \eqno(9b) $$
The value of $x_s$ is in windows $10^{-7}<x_s<10^{-4}$ and $1<Q^2<10 GeV^2$ for the middle and heavy nuclear target.
The results show that the antishadowing effect impedes the shift of $x_s$ toward smaller value with increasing $Q^2$ in the 
range of $Q^2<2 GeV^2$. We can also find that the altitude of the plateau almost grows linearly with increasing 
$Q^2$ in Fig. 16. This consists with the geometric scaling [3,4].

   The values of $Q^2_R$ for the nuclear gluon distribution are plotted in Fig. 7 (dashed and dotted curves) using Eq. (8).
The behavior of curves near $Q^2_R(x)\simeq 2 GeV^2$ is due to the antishadowing corrections, which 
locally raise the gluon distribution. 

	Since the MD-DGLAP equation is based on the collinear factorization scheme and its predictions for the parton
distributions are universal and independent of the concrete process, the above mentioned saturation
phenomena can be checked up
in ultra-relativistic heavy ion collisions
such as the rapidity distribution and centrality dependence of particle production.

    However, we still have not found the saturation solution $Q^2_s$ of Eq. (7) in the available nuclear target.
Obviously, this conclusion is different from
the work [9], which uses the GLR-MQ equation to backward evolve the parton distributions from the input distributions of
a free proton at very higher $Q^2$, where the nonlinear terms are negligible. They give $Q^2_s\sim 3-20 GeV^2$
at $10^{-5}<x<10^{-2}$ for Pb($A=208$)[9]. One can understand the above mentioned
difference as follows. Comparing the solid curves of Figs. 3 and 12, we find the gluon density in the nucleus at higher
$Q^2$ is much smaller than that in the proton even if the recombination terms in the evolution equation do not play
a role at such higher $Q^2$. The reason is that the parton distributions at higher $Q^2$ always remember the recombination
effects in their evolution process. Therefore, a larger, but not true, input distribution in the nuclear target
predicted a stronger recombination effect in work [9], which may balance the parton splitting effect in the evolution,
and give a solution in Eq. (7).

	Now we consider the corrections of the non-perturbative nuclear effects in the nuclear input parton distributions.
The fact that the structure functions of bound and free nucleons are not equal has been discovered long ago
and it is called the EMC effect [19]. However, its dynamics is still an open problem since the several mechanisms
dominate the EMC effect.
Although there are some models to describe the non-perturbative nuclear shadowing relating to the GRV-input distributions
[20,21], they still have a larger uncertainty, 
particularly
in the nuclear gluon distribution. In this case, we take following simplified factors $R_S$ and $R_G$ 
to describe the contributions of the non-perturbative nuclear shadowing like Ref. [22],

$$S_A(x,Q^2_0)=R_S(x,Q^2_0,A)S(x,Q^2_0),\eqno(10) $$
for sea quarks and

$$G_A(x,Q^2_0)=R_G(x,Q^2_0,A)G(x,Q^2_0),\eqno(11) $$
for gluon, where

$$R_S(x,Q^2_0,A)=\left\{\begin{array}{ll}
                     1,& x_n<x<1\\
                     1-K_S(A^{1/3}-1)\frac{x_n-x}{x_n-x_A},& x_A\leq x\leq x_n\\
                     1-K_S(A^{1/3}-1), & 0<x<x_A,\\
                    \end{array}
                     \right. , \eqno(12)$$ 
and
$$R_G(x,Q^2_0,A)=\left\{\begin{array}{ll}
                     1,& x_n<x<1\\
                     1-K_G(A^{1/3}-1)\frac{x_n-x}{x_n-x_A},& x_A\leq x\leq x_n\\
                     1-K_G(A^{1/3}-1), & 0<x<x_A,\\
                    \end{array}
                     \right.,  \eqno(13)$$ 
in which $x_n=0.05$, $x_A=0.017A^{-1/3}$, and $K_S=0.09$ are parameterized using the data 
about the EMC effect [19]; while we take $K_G=K_S$ for the moment.
In the meantime, we neglect the
EMC effects in $x>0.05$ since we focus the behavior of parton distributions at very small $x$ limit. 
Using Eqs. (10)-(13) as the input distributions we evolve the MD-DGLAP equation. Our results are shown 
in Figs. 17-19 for Ca($A=40$) and Pb($A=208$). Comparing these results with Eqs. 7-9,
we find that the differences between them are small.

	Since we have not enough data to fix the value of $K_G$, we take the parton distributions for
Pb($A=208$) as an example and change $K_G$ from $0.18$ to $0$ 
to test the sensitivity to such choices. We find that the results are insensitive to the 
above mentioned change since the gluon distribution Eq. (3) in $x<0.05$ is very small.

	Now we can conclude that the saturation- or partial saturation
phenomena in the small $x$ behavior of parton distributions are dominated by QCD dynamics rather 
than non-perturbative nuclear shadowing corrections in the input distributions.

\newpage
\begin{center}
\section{Discussions and summary}
\end{center}

	Let us discuss the antishadowing contributions.  Part of momentum is lost 
due to the negative shadowing terms in the GLR-MQ equation. Most works do not make an
attempt to correct the evolution equation from the momentum conservation since the lost
momentum is only a few percent of the total momentum. However, as one of us (WZ) has pointed
out that the antishadowing contributions, which balance the lost momentum in the shadowing 
effect can not be neglected [23]. For illustrating this point, we set the antishadowing terms to zero in the 
MD-DGLAP equation (1) and use it to evolve the parton distributions. In this case, we re-fit the input
distributions (4) and take $K=0.000285$ using the HERA data [13]. New parton distributions
in proton, Ca($A=40$) and Pb($A=208$) are plotted by dotted curves in Figs. 3-4, 8-9 and 12-13, respectively. 
There are following two distinguishing features as compared with solid curves:

 (1) The results show that the parton distributions with antishadowing effect are lower than that without
antishadowing effect.
One can understand the above-mentioned results as follows.
The main HERA data, which are used to fit the nonlinear parameter $K$ have shown the steeper form
at $Q^2> 1GeV^2$ and $x>10^{-4}$,
where the antishadowing correction is larger. Therefore, to fit the same HERA data, the value of $K$ in the 
evolution 
equation with the antishadowing terms should be larger than that 
in the evolution equation without the antishadowing terms, since in the former case an obvious antishadowing 
effect partly cancels
the shadowing effect. On the other hand, the net screening effect flats the gluon distribution when $x$ enters 
into a smaller $x$ region and therefore weakens the antishadowing effect. Therefore, a more stronger net screening effect
appeared in Eq. (1). Comparing the solid curves with the dotted curves in Figs. 3 and 4, one can understand that the 
antishadowing corrections make the growth of the solid curves 
slower than that of the dotted curves.

	The above mentioned differences between the effects with antishadowing and without antishadowing also appear 
in the predictions of the MD-DGLAP equation (with antishadowing) and the GLR-MQ equation (without antishadowing). 
Using the same program as we used in the
MD-DGLAP equation, we fit the relating parameters in the GLR-MQ equation when using the GRV-like input
parton distributions. Then we predict the parton distributions in proton and Pb($A=208$) (see Figs. 20
and 21). The results show that the screening effect in the GLR-MQ equation is weaker than the net-screening
effect in the MD-DGLAP equation if we use similar input parton distributions. Although the plateau
has appeared in the parton distributions of Pb($A=208$), we still have not found saturation solution
$Q^2_s$ in Eq. (6) for the GLR-MQ equation. 

	(2) An other interesting difference between the predictions of the equations without and with antishadowing 
corrections is the position of $x_s$. In the former case $x_s$ always moves toward smaller value with increasing $Q^2$
(note that $x_s$ runs out the diagrams, for example, at $Q^2\geq 2 GeV^2$ in Fig. 12),
while in the later case this shift is impeded by the antishadowing effect in the range of lower $Q^2$ ($Q^2<2 GeV^2$). 
This phenomenon relates to Fig. 7, where the line has a corner near $Q^2_R(x)\simeq 2 GeV^2$ due to the 
relative stronger antishadowing effect at lower $Q^2$ locally raises the gluon distribution.

	As we know that the nuclear shadowing has different manifestations: the suppression in the usual or 
unintegrated parton distributions and their evolutions, the reduction of the structure functions and 
the screening effect in the cross sections. There are many different kinds of model to study the above 
mentioned shadowing phenomena.
In this work we do not try to compare the modified DGLAP equations with other versions 
of the saturation model such as the JIMWLK equation [4]. The reason is that they have different research 
subjects: the former case discusses the
shadowing in the parton distributions, which are independent of the process, while the 
later case treats the unintegrated parton distributions and the cross sections, which 
are process-dependent. 

	Finally, we take the LLA($Q^2$) in this work. The contributions beyond the LLA($Q^2$) are necessary for the improvement
of predictions in the range of very lower $Q^2$ ($Q^2< 1 GeV^2$). However, it relates to the further study
including the evolution dynamics at higher order.

	In summary, the parton distributions in small $x$ region are numerically
predicted by using a modified DGLAP equation with the GRV-like input distributions.
We show that the gluon recombination at twist-4 level obviously 
suppresses the rapid growth of parton densities with $x$ decrease.
The growth of predicted parton distributions in proton towards small $x$ is slower than $\ln (1/x)$ in $x<10^{-6}$. 
In particular, a plateau is formed in the parton distributions of the nuclear target at small $x$ limit 
and $1<Q^2<10 GeV^2$.  
The altitude of the plateau almost grows linearly with $Q^2$ increasing, i.e., the geometrical scaling.
Thus, the parton distributions in proton and nucleus unitarize, and the Froissart
boundary is not violated in the asymptotic regime of high density QCD. 
The saturation (or partial saturation) phenomena appear before the saturation scale $Q^2_s$, 
where the gluon recombination correction fully balance the parton splitting effects. 
The predicted saturation phenomena can be checked up  
in ultra-relativistic heavy ion collisions
such as the rapidity distribution and centrality dependence of particle production.

\vspace{0.3cm}

\noindent {\bf Acknowledgments}:

  We would like to thank A. De Roeck, M. Gl\"{u}ck and E.Reya for helpful
    discussions. It is a pleasure for us to gratefully acknowledge
    the referees' suggestions that lead to the present form of our
    manuscript. This work was supported by following National Natural Science Foundations of China 10075020, 90103013,
10135060, 10205004 and 10175074.

 \newpage

Figure Captions

Fig. 1 The fits of the computed $F_2(x,Q^2)$ in proton by the MD-DGLAP equation (solid curves) to (a) H1-
and (b) ZEUS-data.
The dashed curves are the DGLAP equation results from GRV98LO.

Fig. 2 The fits of the computed $dF_2(x,Q^2)/d\ln Q^2$ in proton by the MD-DGLAP equation (solid curves) to the H1 data,.
The dashed curves are the DGLAP equation results from GRV98LO.

Fig. 3 The predictions for the gluon distribution function in proton. The solid (dotted)
curves are the results of the MD-DGLAP equation with (without) antishadowing corrections; The dashed curves
are the results of the DGLAP equation with the GRV98LO. Notice that the solid and dashed curves have opposite concavities.

Fig. 4 As Fig. 3 but for the sea-quark distribution function in proton.

Fig. 5 The $Q^2$-dependence of the gluon distribution function in proton at fixed values of $x$, the 
solid (dotted) curves correspond to with (without) antishadowing contributions. The dashed curves are the results of
the DGLAP equation.

Fig. 6 As Fig. 5 but for the sea-quark distribution function in proton.

Fig. 7 The gluon recombination scale $Q^2_R(x)$ in the MD-DGLAP equation for proton (solid curve), Ca($A=40$) 
(dotted curve) and Pb($A=208$) (dashed curve); where the input distributions are 
Eqs. (3) and (4).

Fig. 8 The predictions for the gluon distribution function in Ca($A=40$). The solid (dotted)
curves are the results of the MD-DGLAP equation with (without) antishadowing contributions; The dashed curves
are the results of the DGLAP equation; where the input distributions are 
Eqs. (3) and (4).

Fig. 9 As Fig. 8 but for the sea-quark distribution function.

Fig.10 The $Q^2$ dependence of the gluon distribution function in Ca($A=40$) at fixed values of $x$, from with 
(solid curves) and without (dotted curves) antishadowing corrections in the present work. The dashed curves are
the results of
the DGLAP equation. Notice the plateau at small $x$; where the input distributions are 
Eqs. (3) and (4).
.

Fig. 11 As Fig. 10 but for the sea-quark distribution function.

Fig. 12 As Fig. 8 but for Pb($A=208$).              

Fig. 13 As Fig. 9 but for Pb($A=208$).             

Fig. 14 As Fig. 10 but for Pb($A=208$).
              
Fig. 15 As Fig. 11 but for Pb($A=208$).              

Fig. 16 Parts of Figs. 14 and 15. The lines show that the altitude of the plateau in Ca($A=40$) and Pb($A=208$) 
almost grows linearly with increasing $Q^2$. 

Fig. 17 The predictions for the gluon distribution function in Ca($A=40$) and Pb($A=208$)
using the MD-GDLAP equation and input distributions
Eqs. (10)-(13).

Fig. 18 The predictions for the quark distribution functions in Ca($A=40$) and Pb($A=208$)
using the MD-GDLAP equation and input distributions
Eqs. (10)-(13).

Fig. 19 As Fig. 7 but the input distributions are taken as Eqs. (10)-(13).

Fig. 20 The predictions for the gluon distribution function using the GLR-MQ evolution equation with
a GRV-like input parton distributions. The solid and dotted
curves are the results in proton and Pb($A=208$).

Fig. 21 As Fig. 20 but for the sea-quark distribution function.

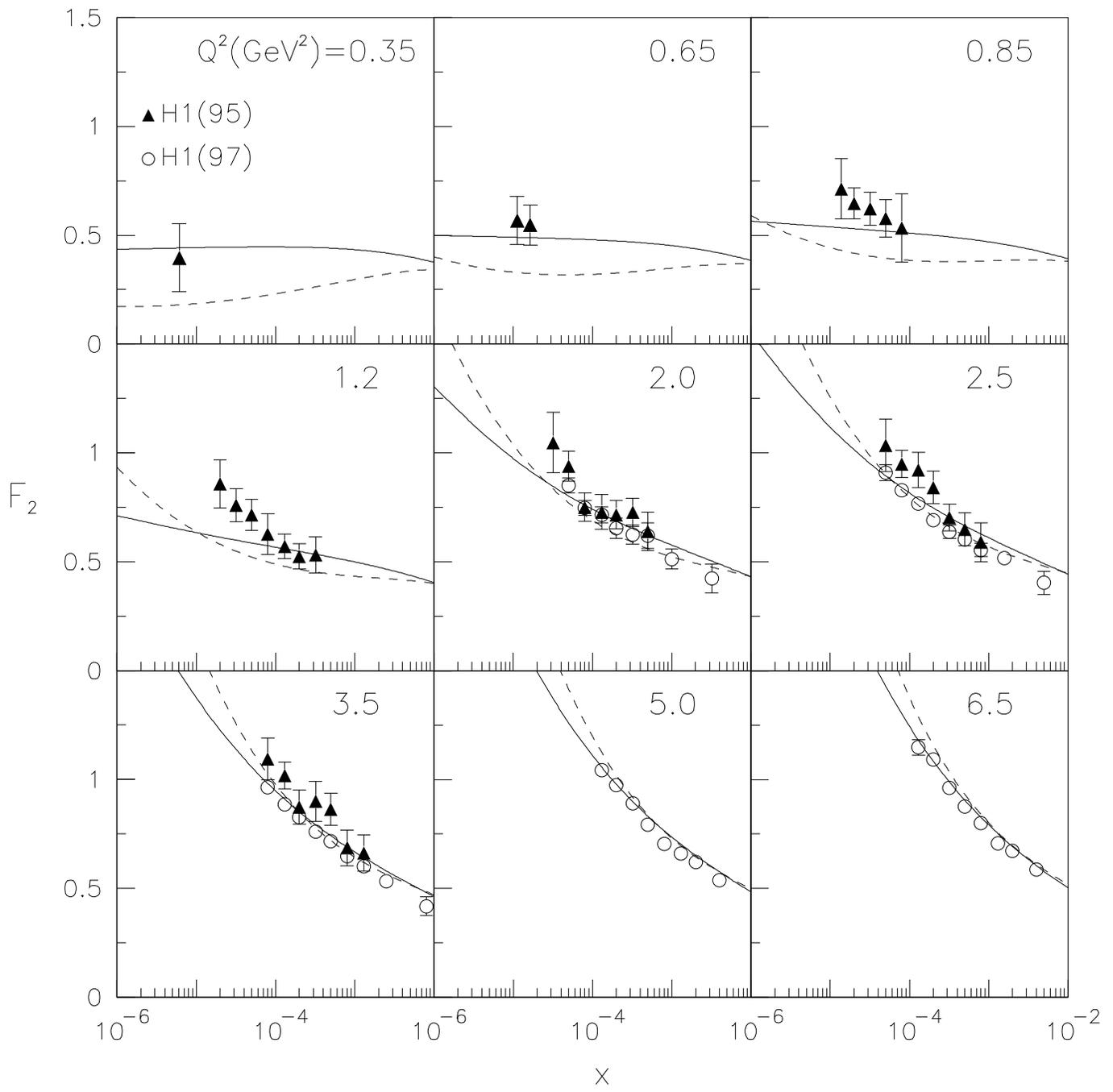
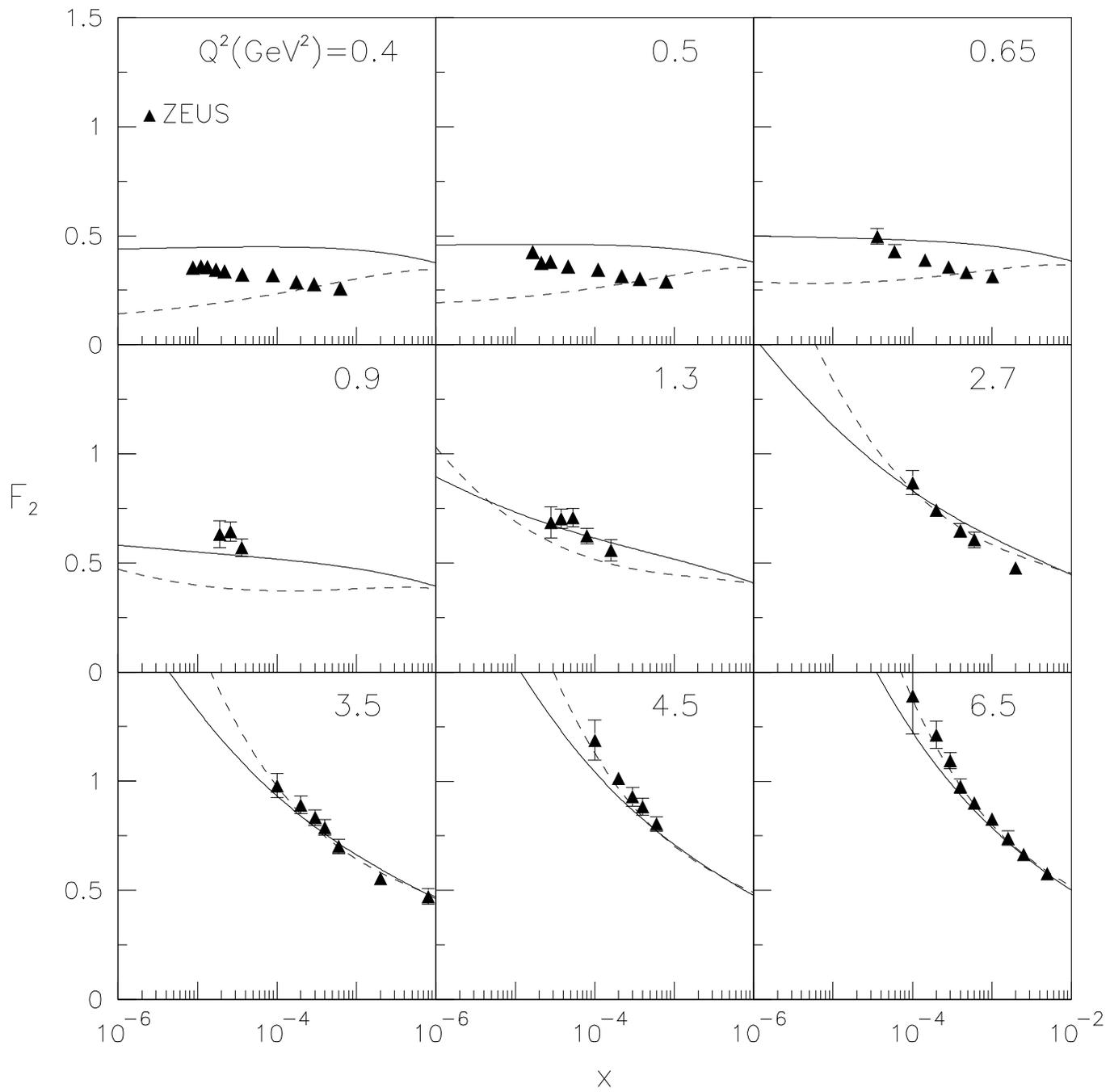
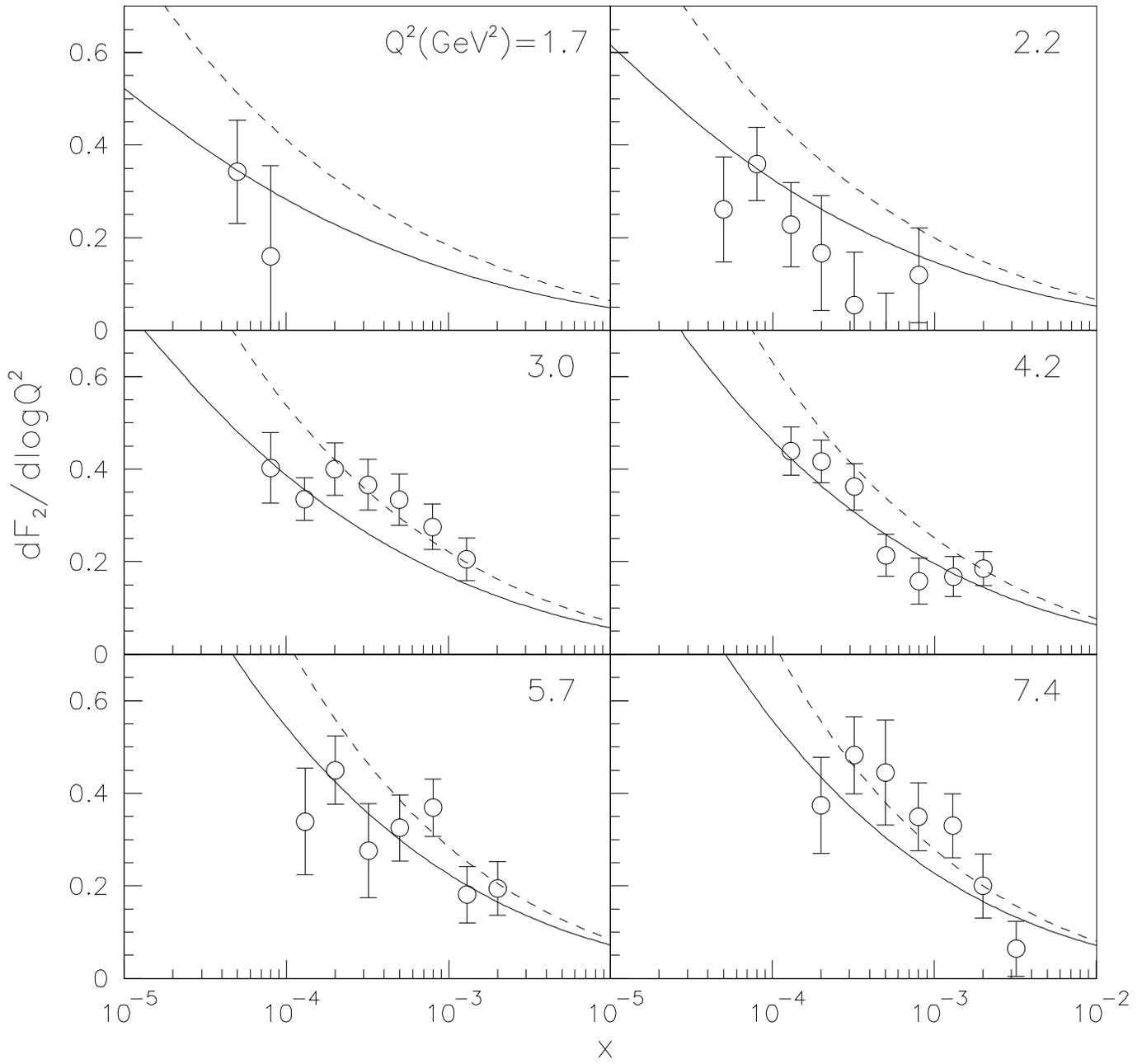
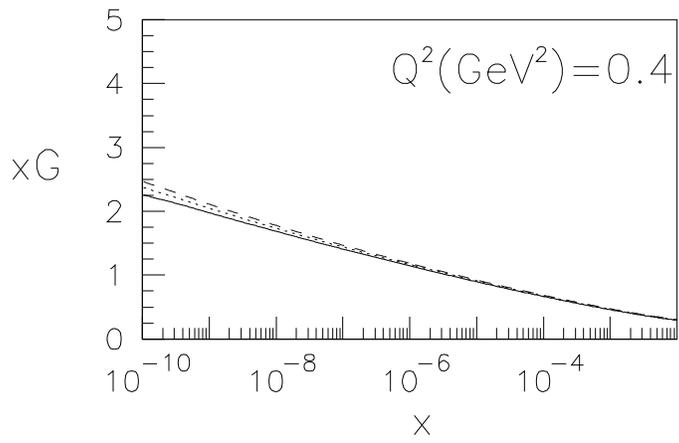
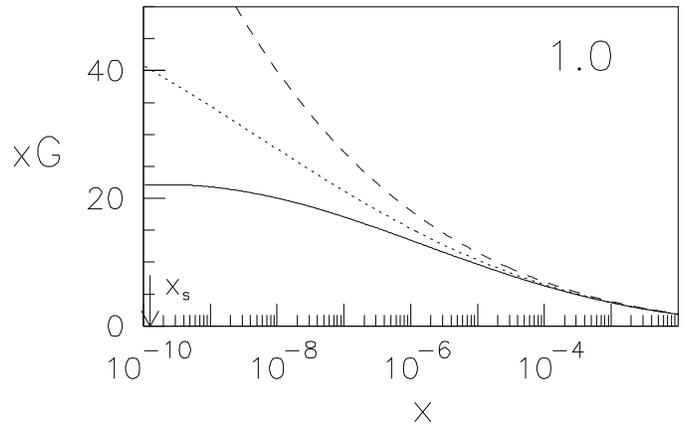
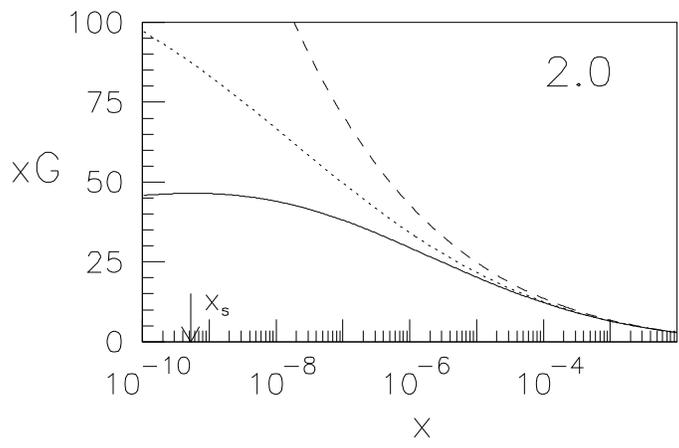
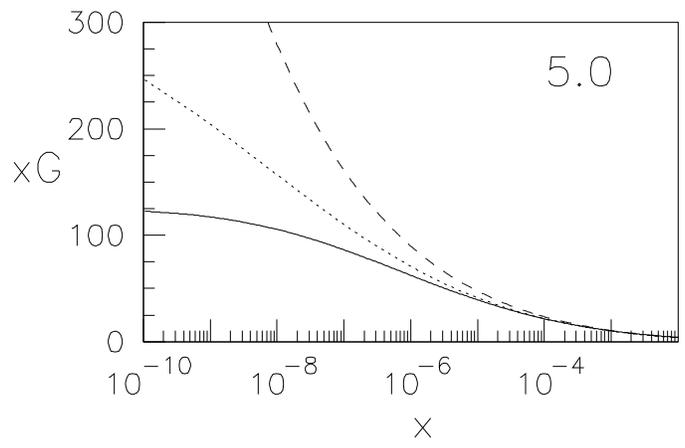
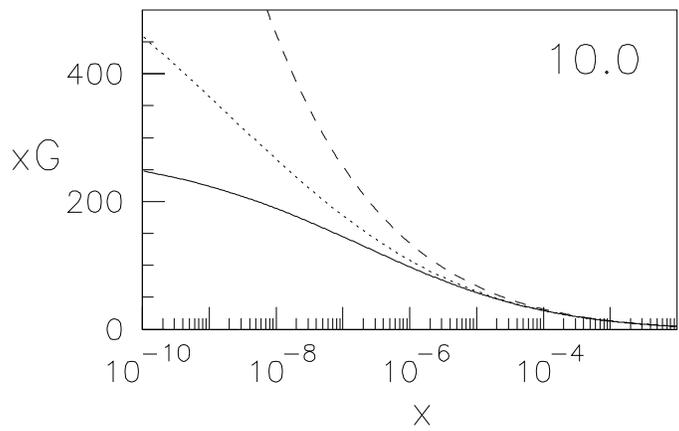
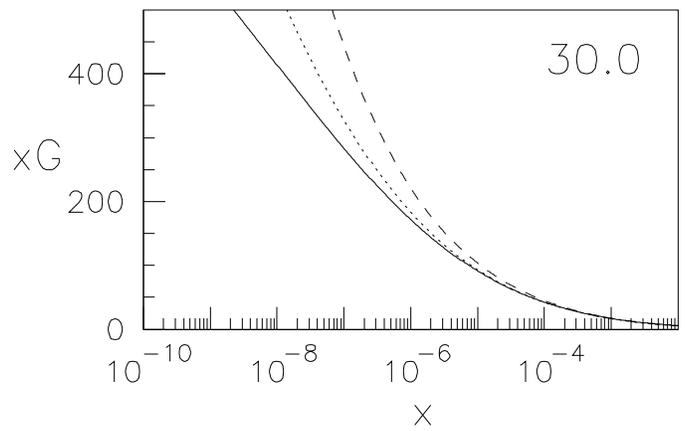
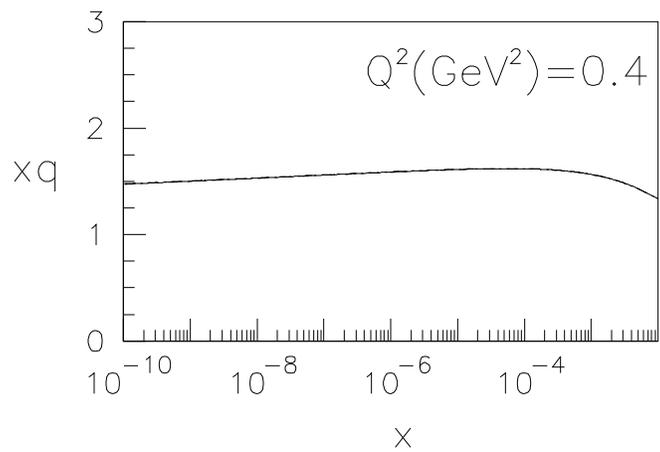
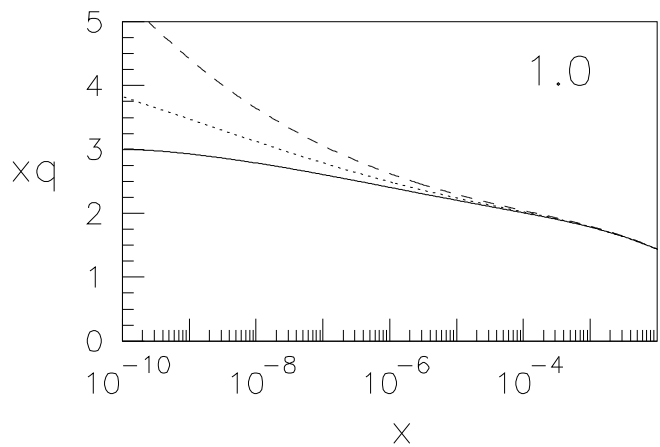
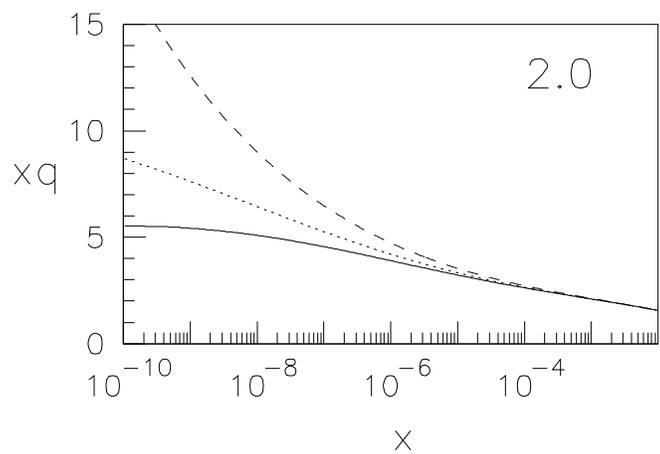
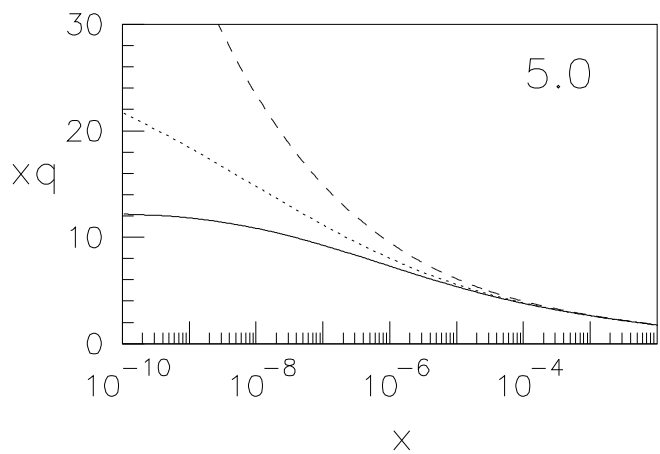
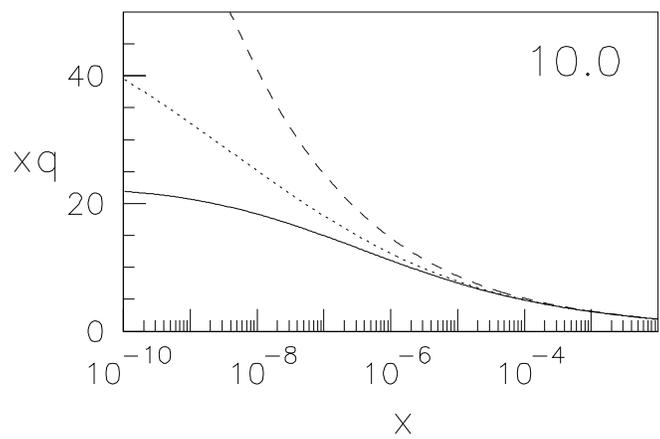
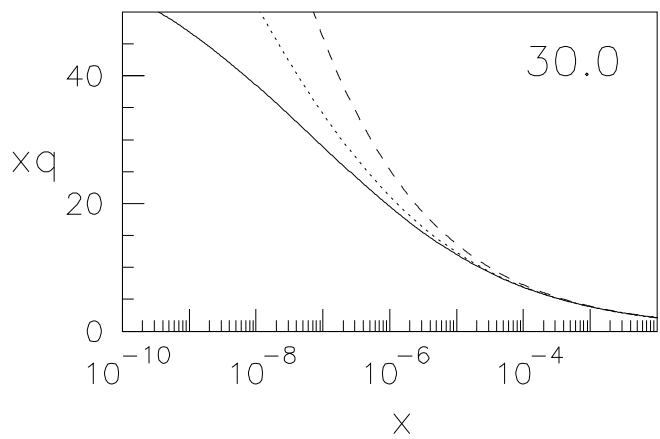
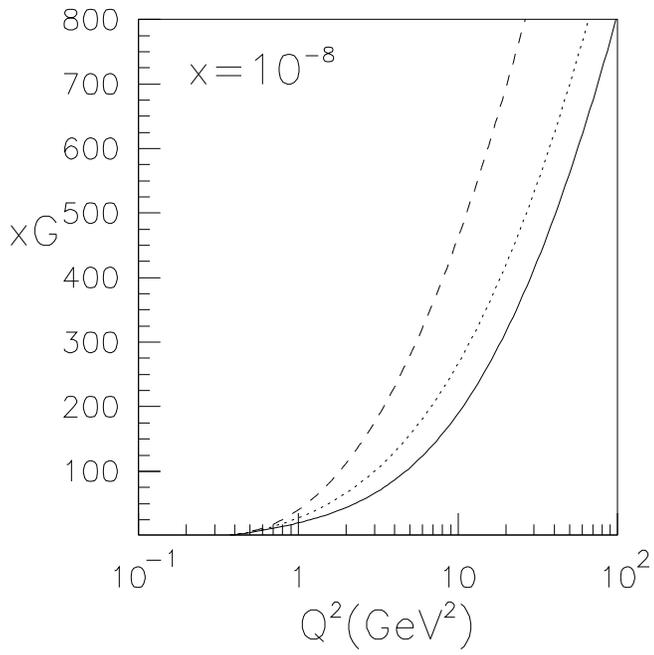
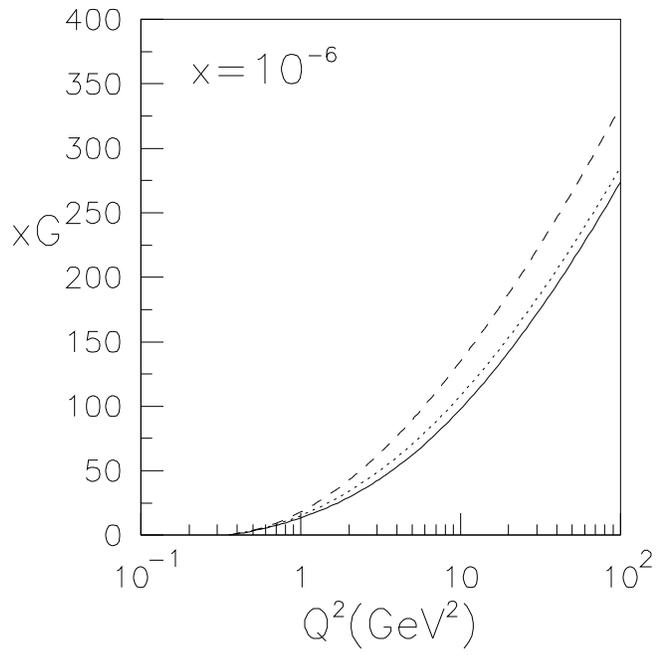
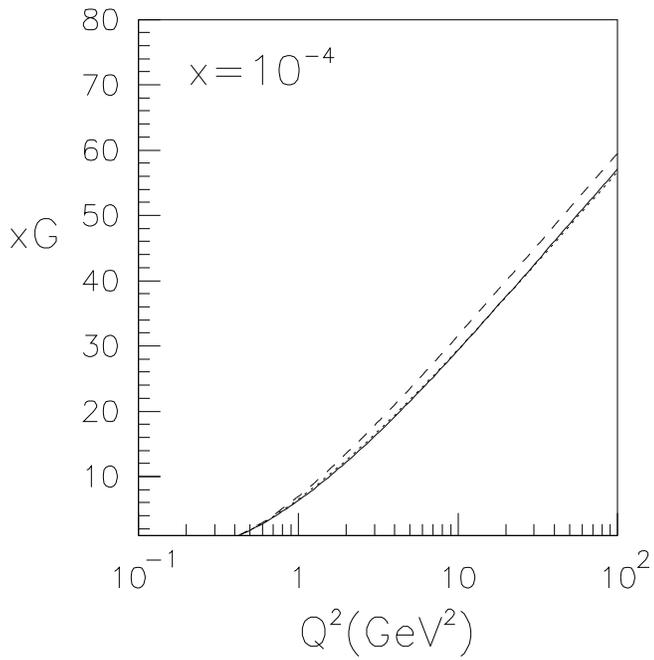
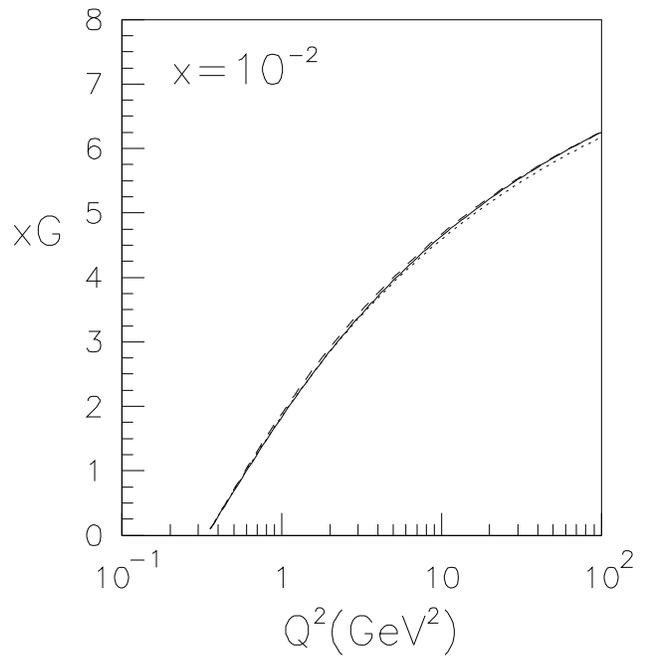
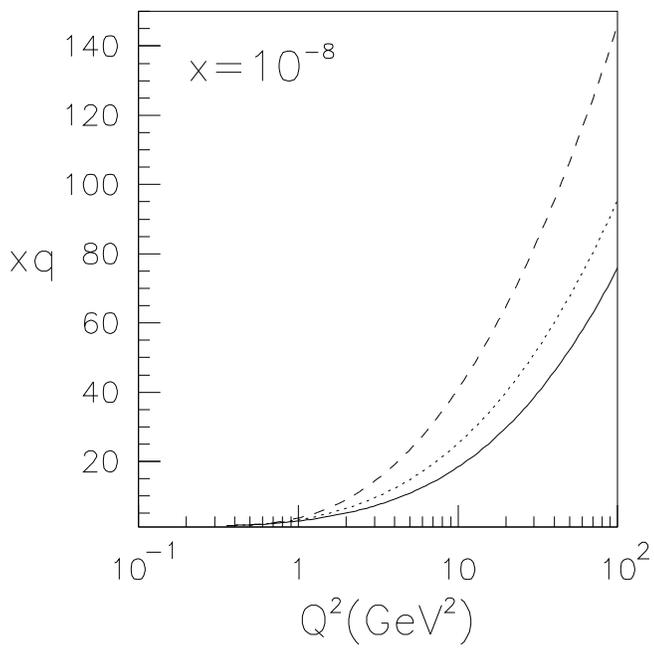
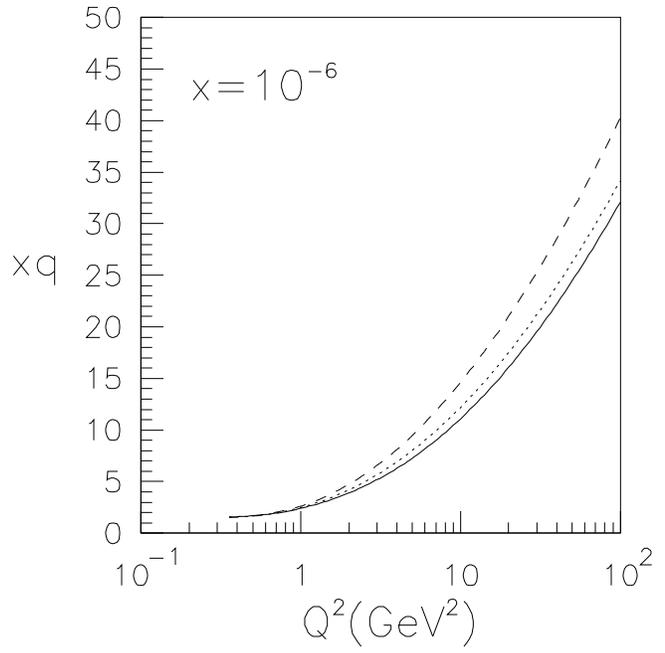
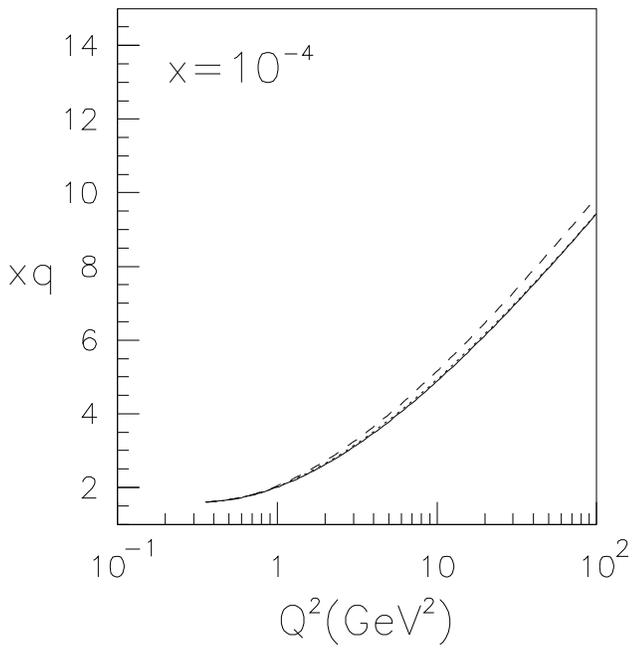
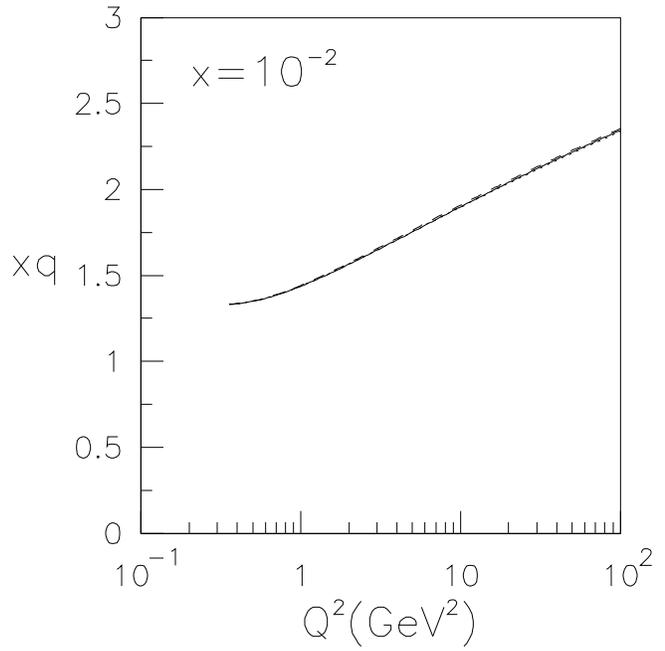
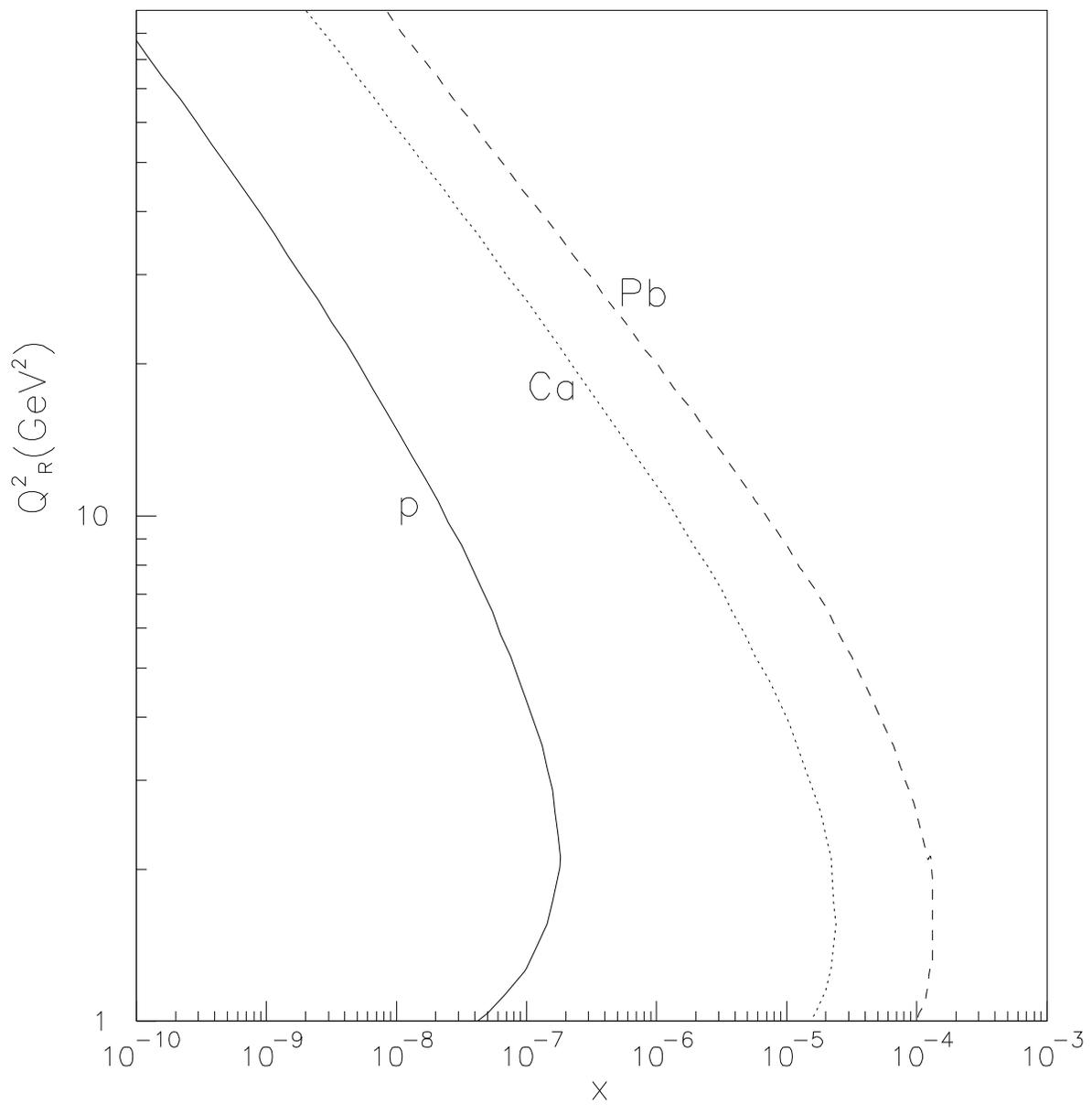
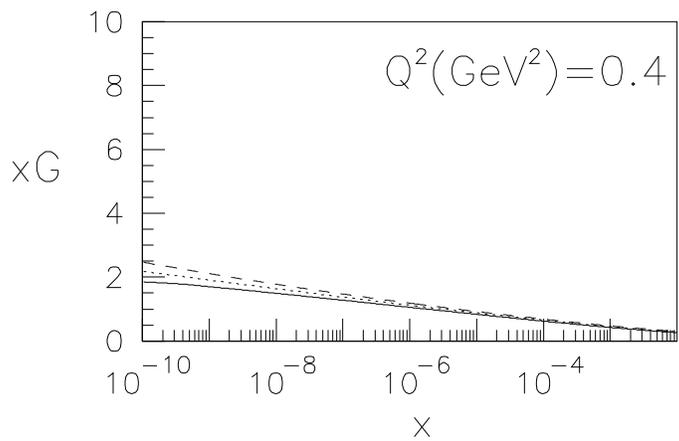
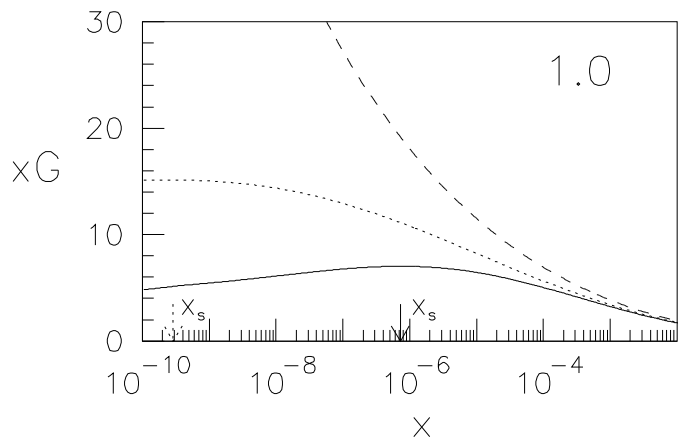
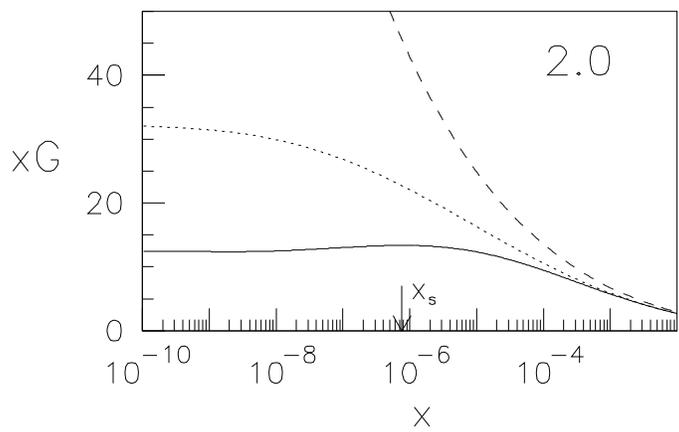
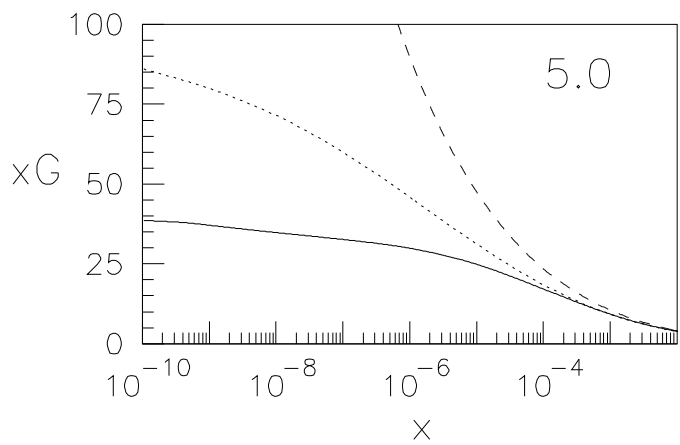
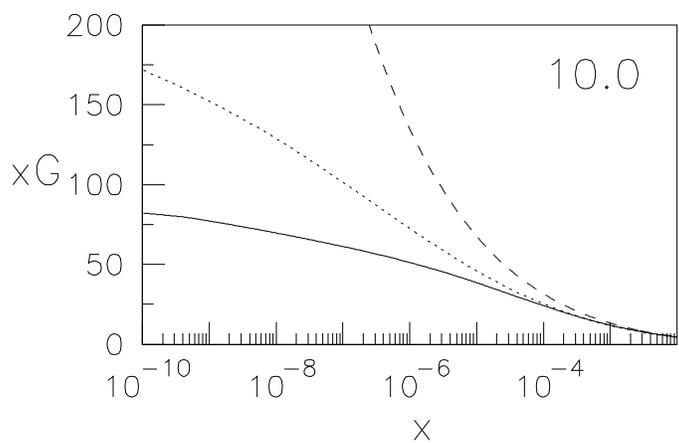
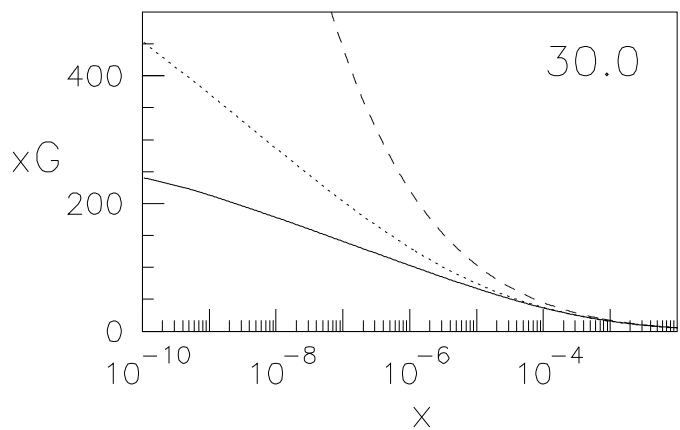
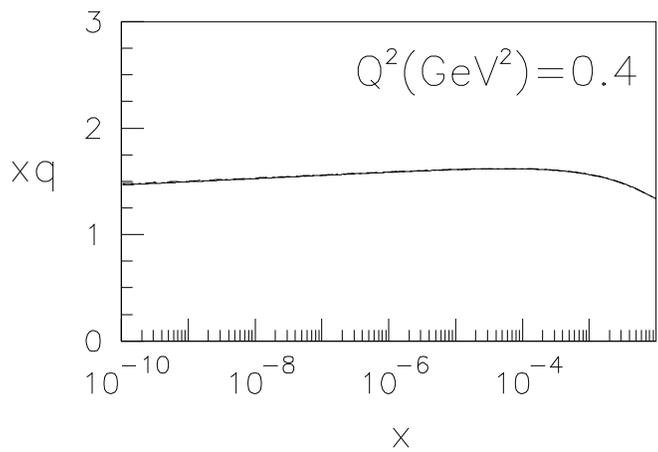
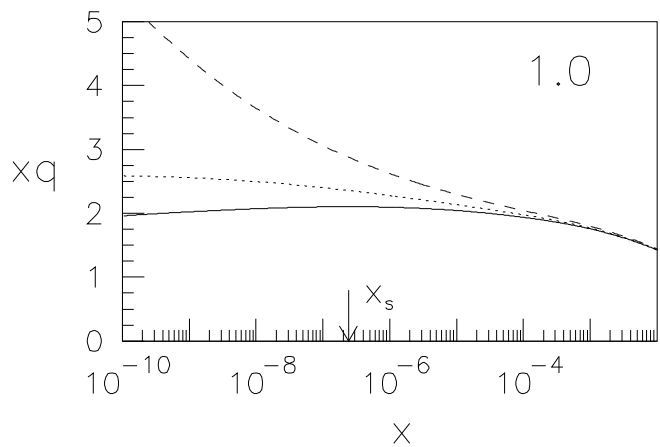
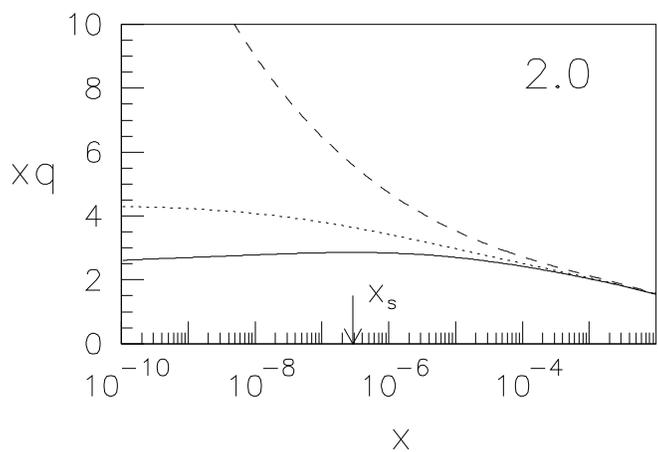
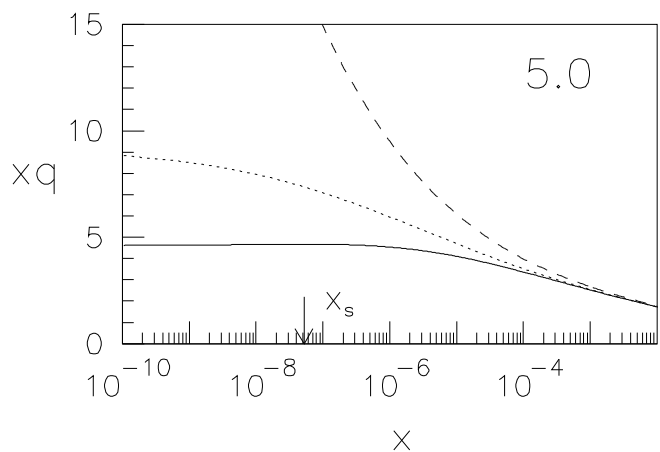
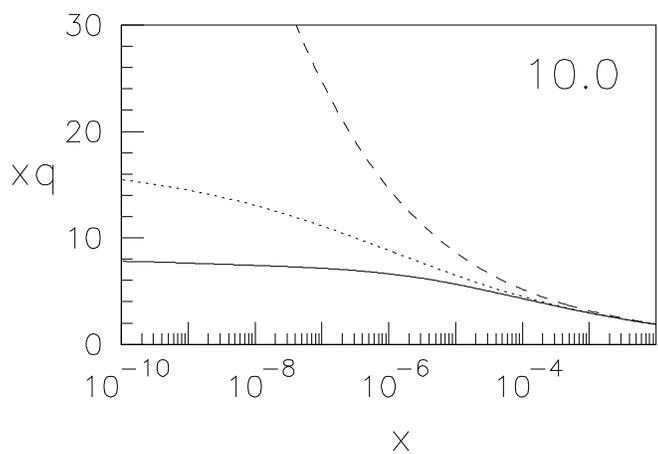
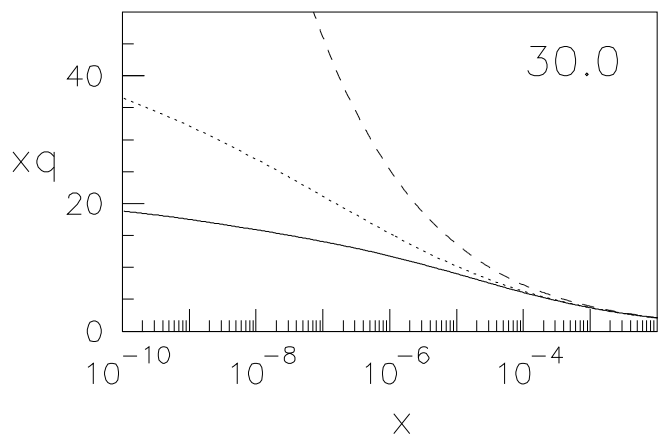
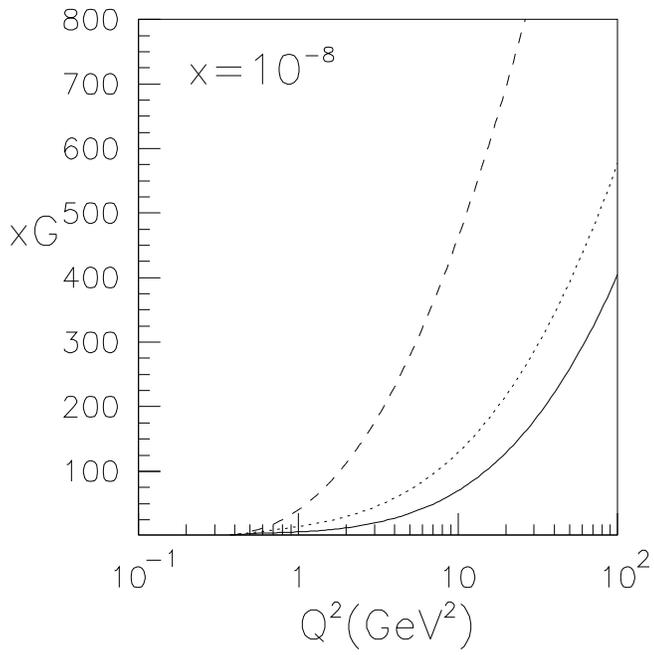
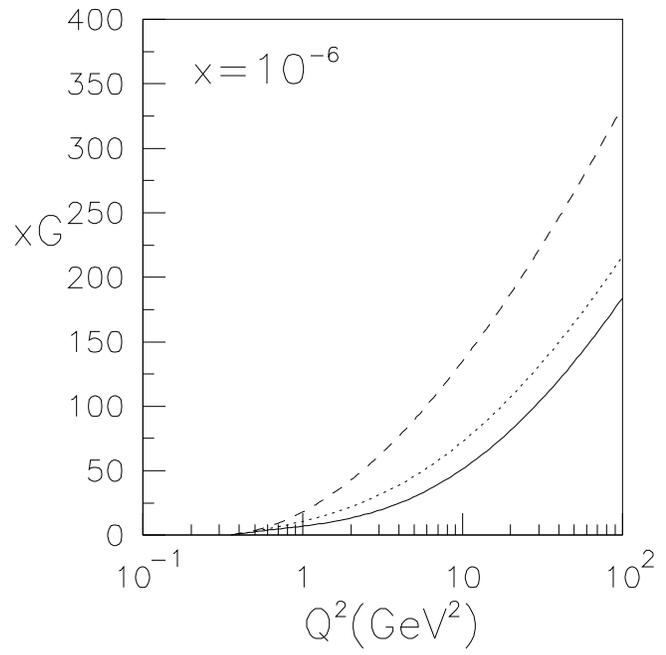
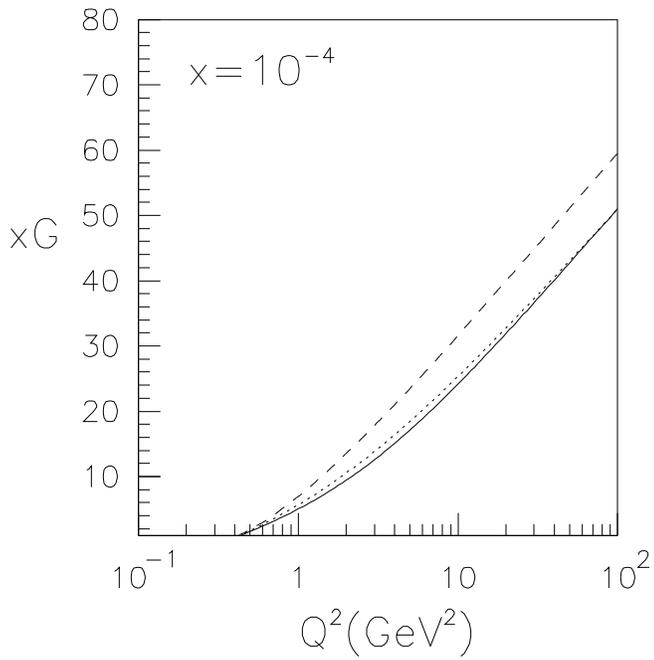
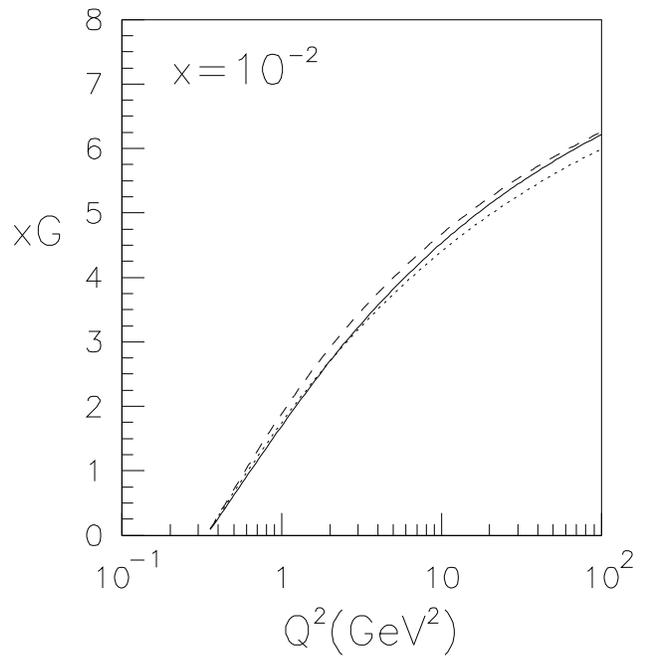
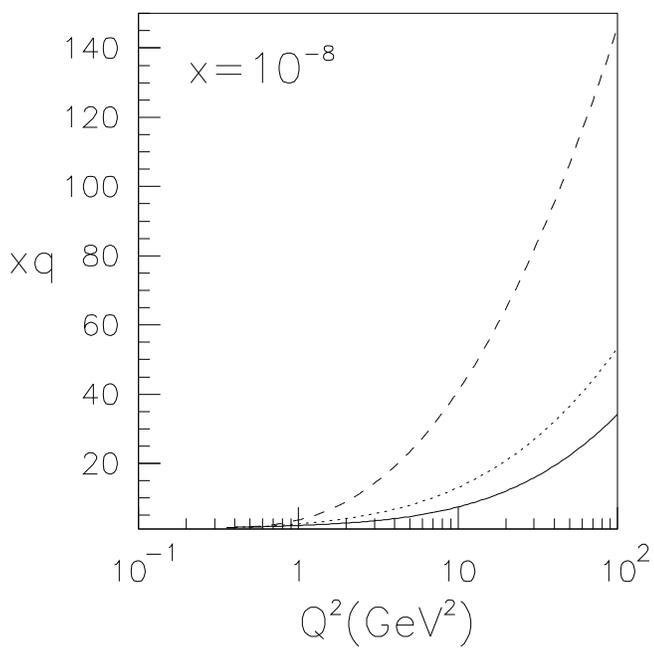
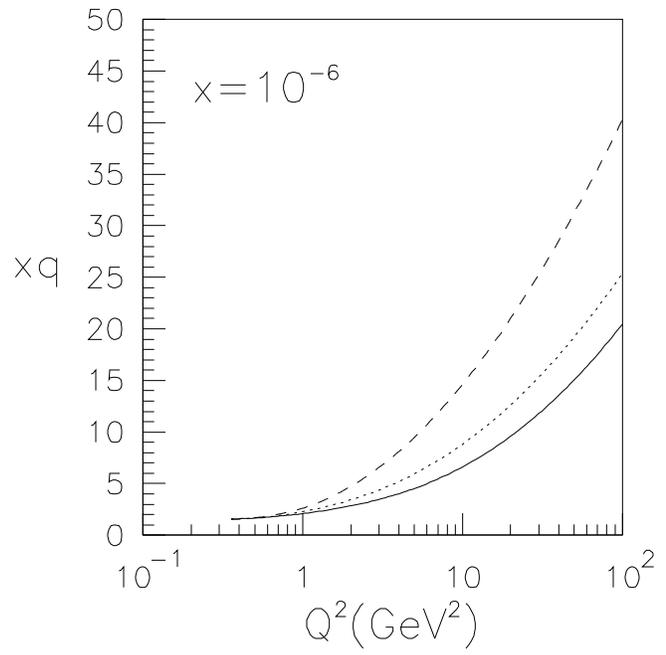
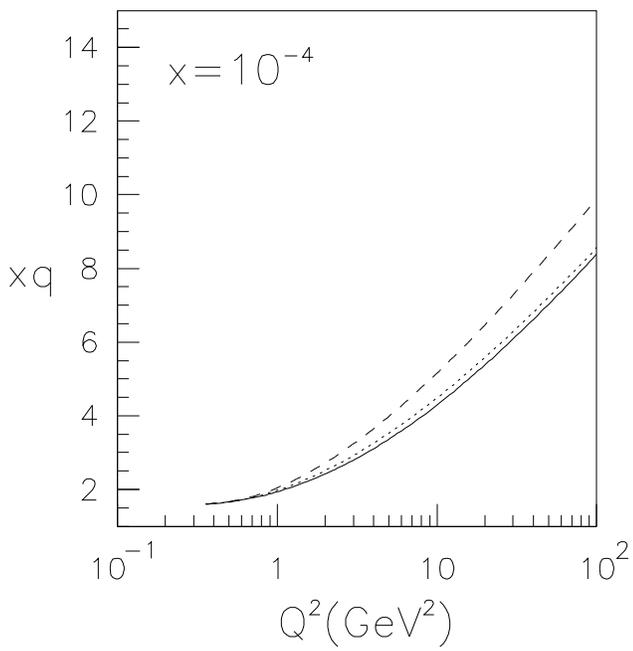
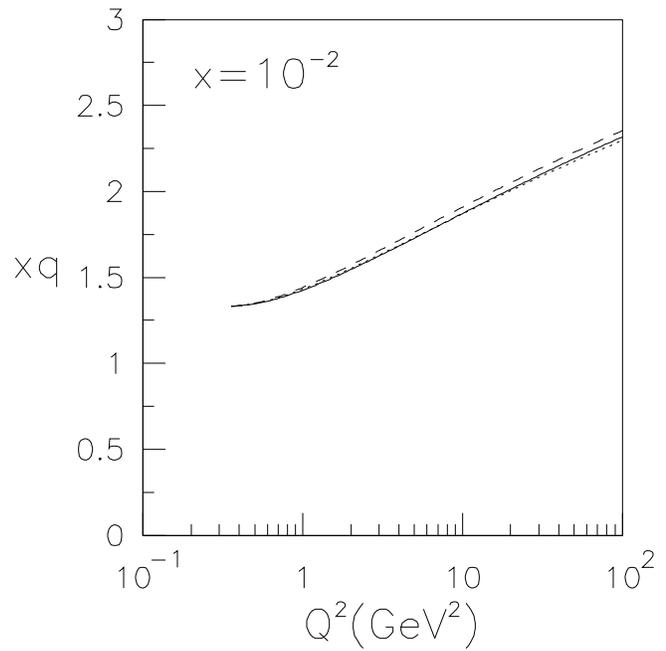
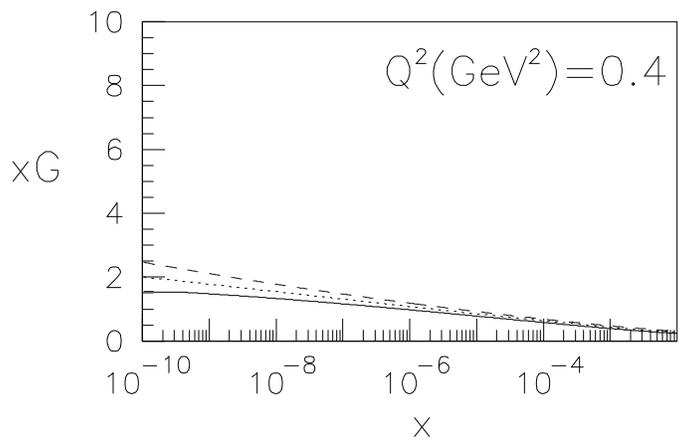
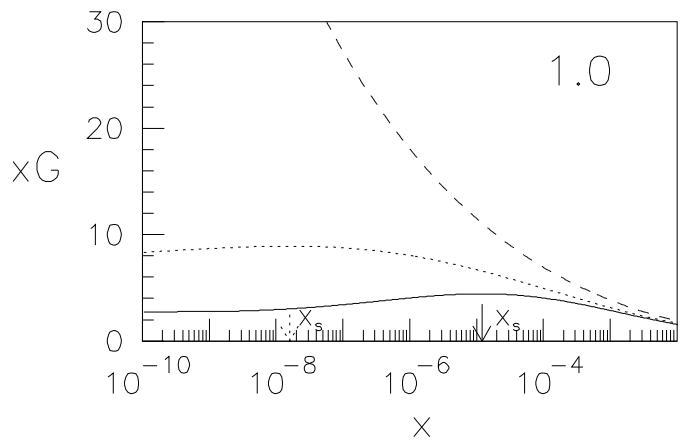
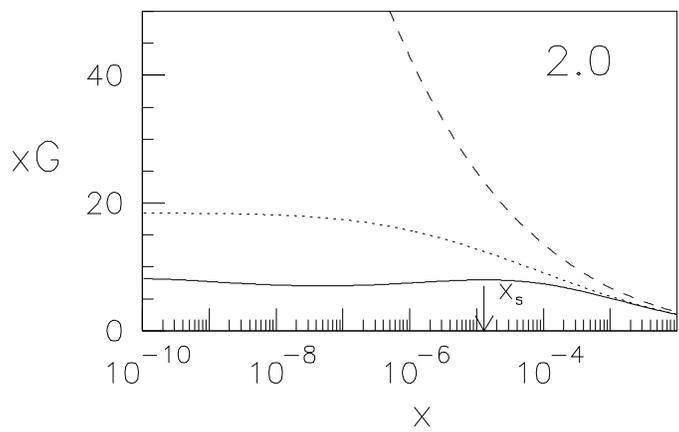
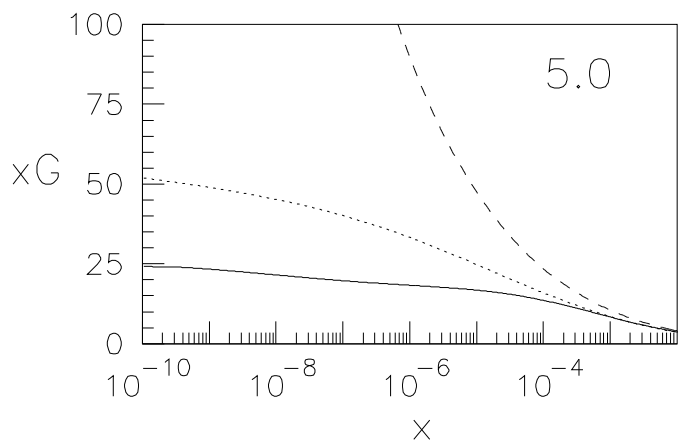
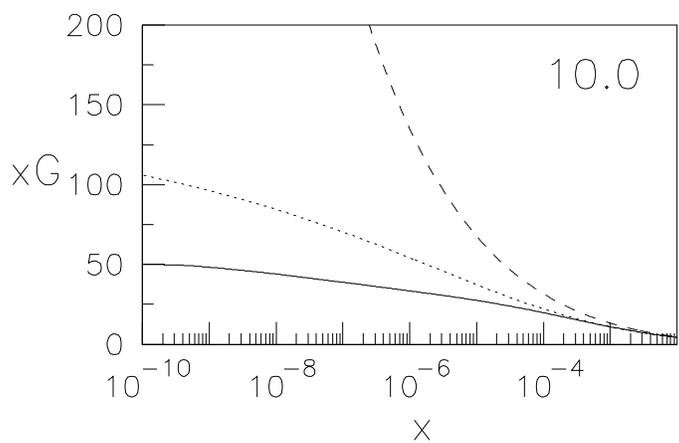
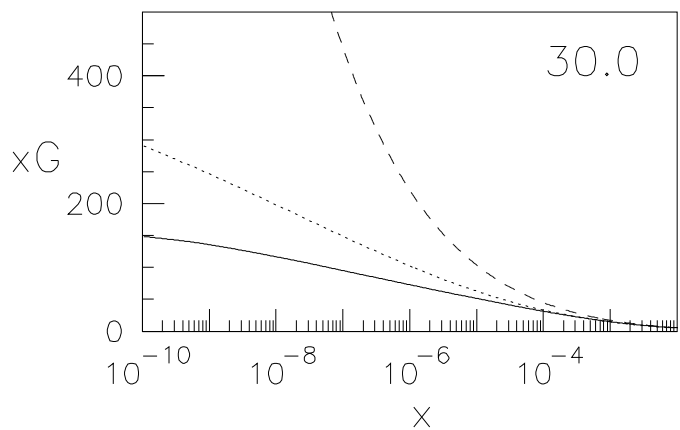
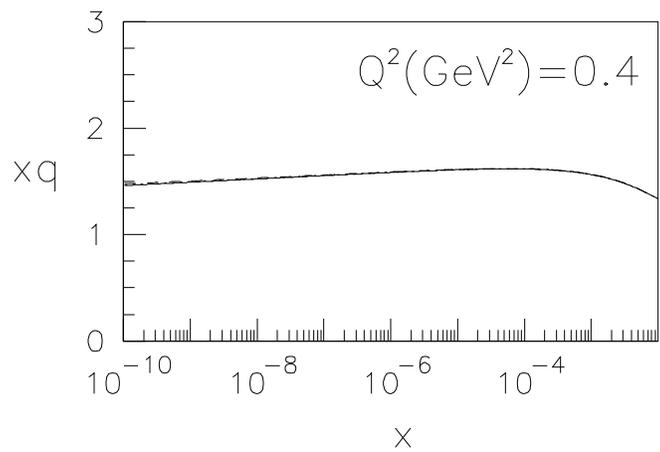
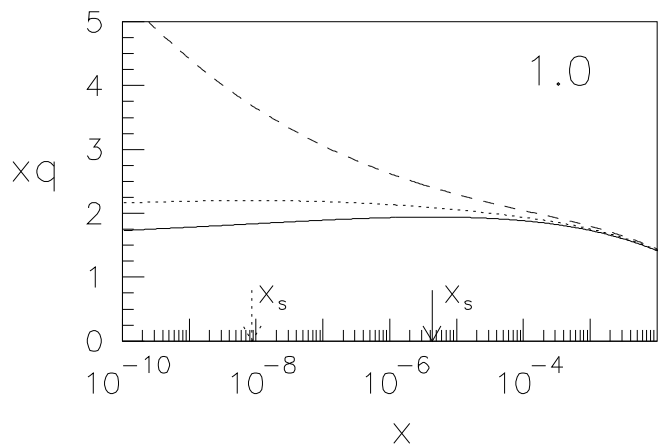
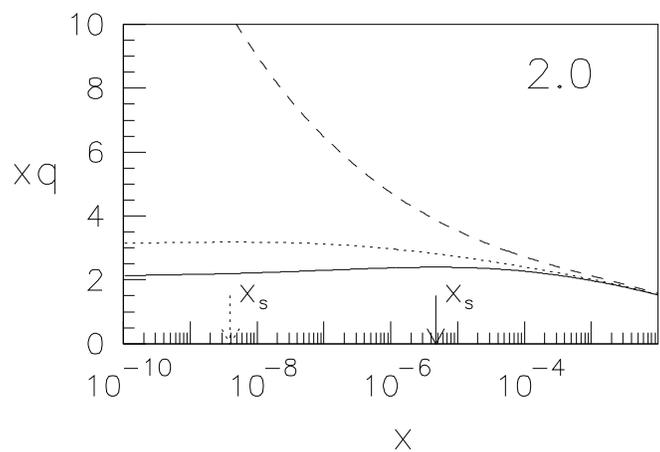
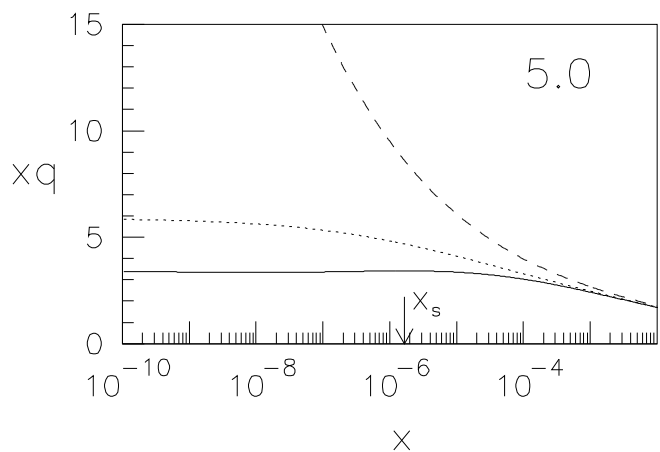
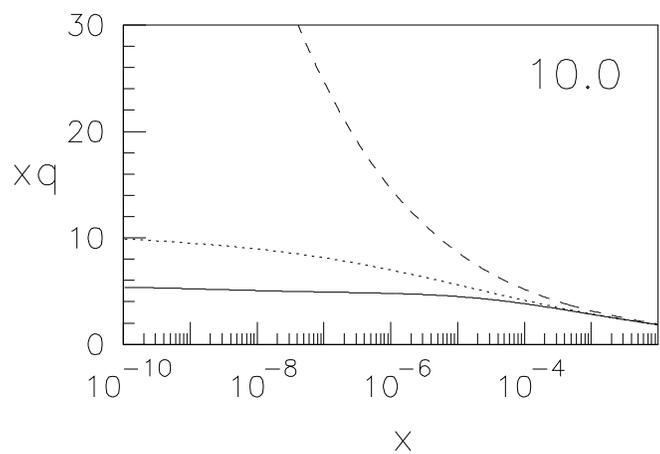
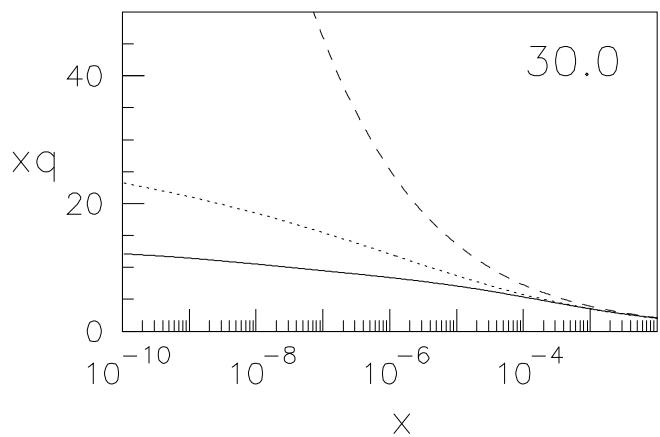
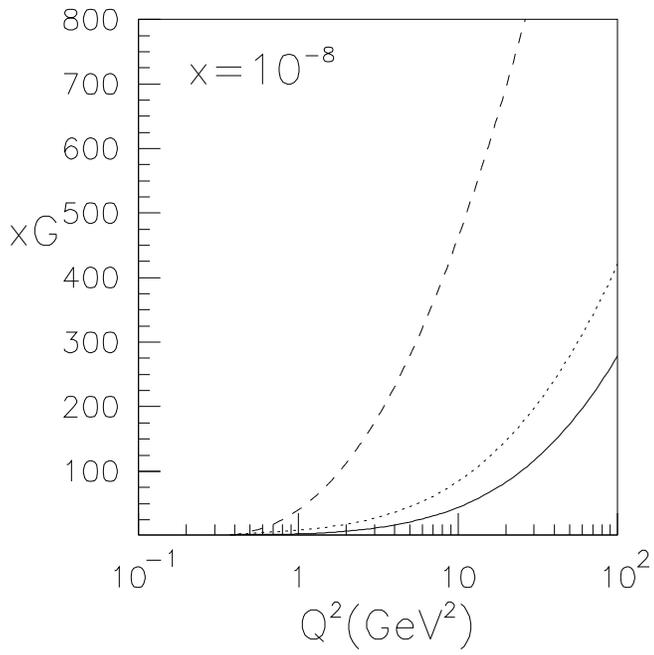
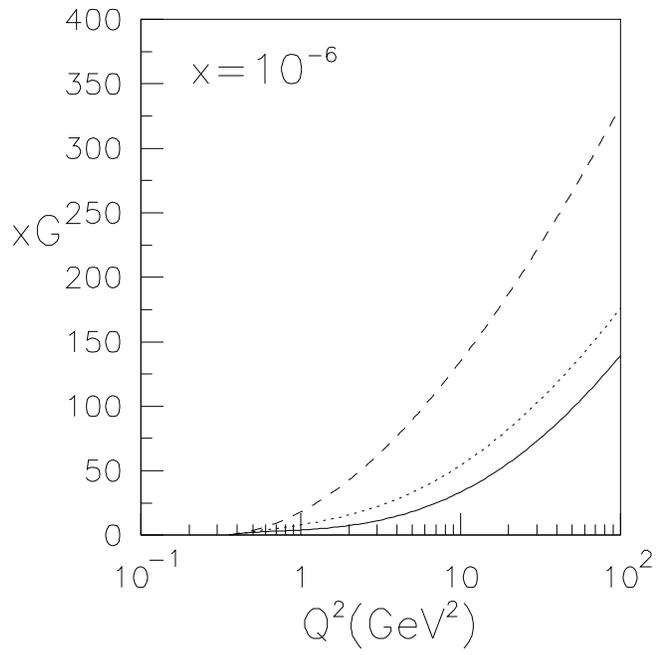
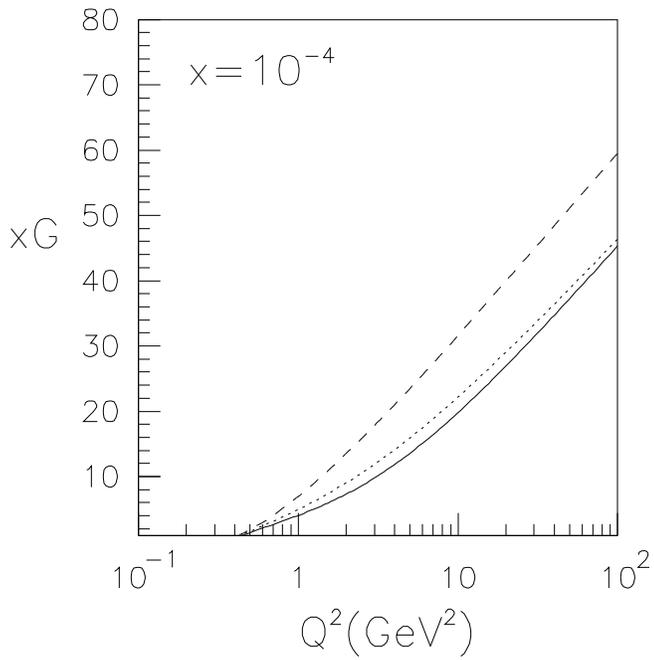
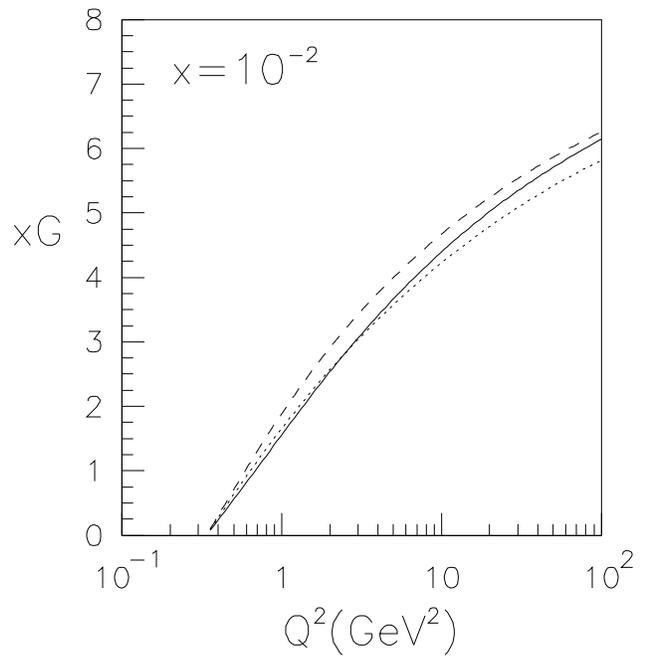
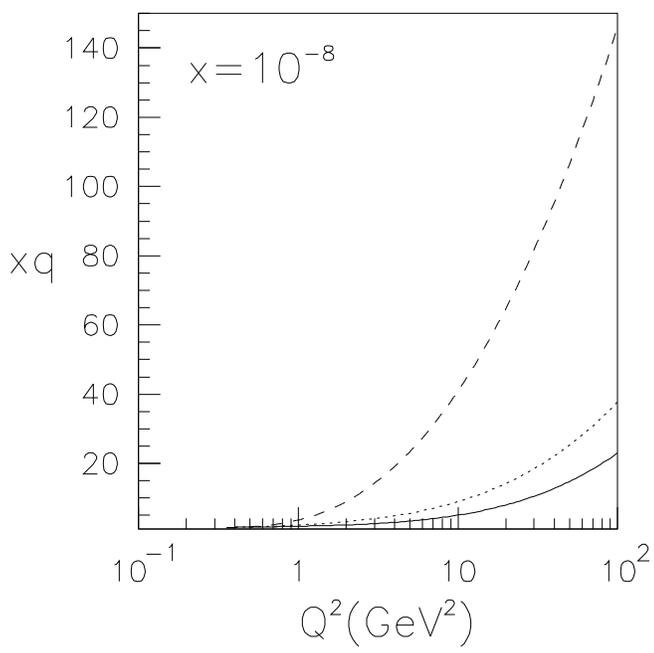
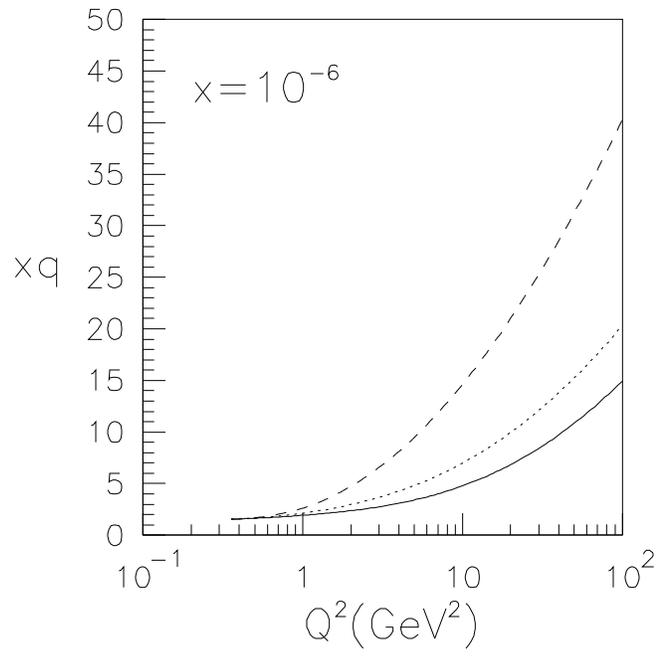
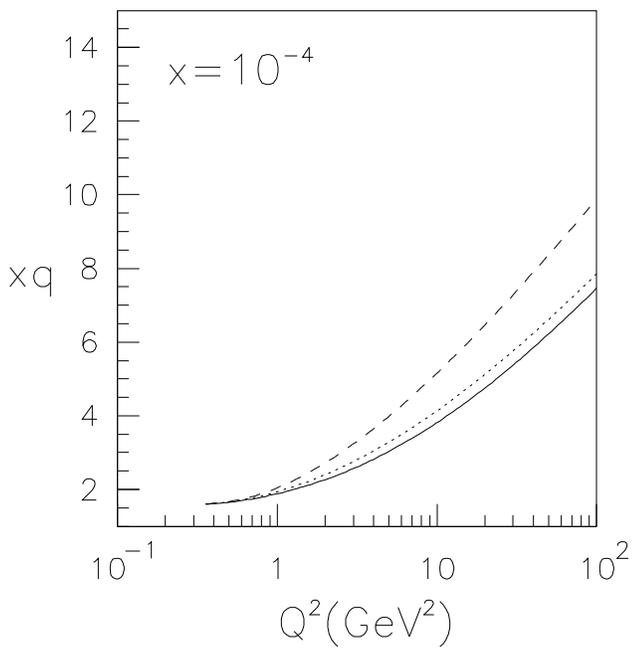
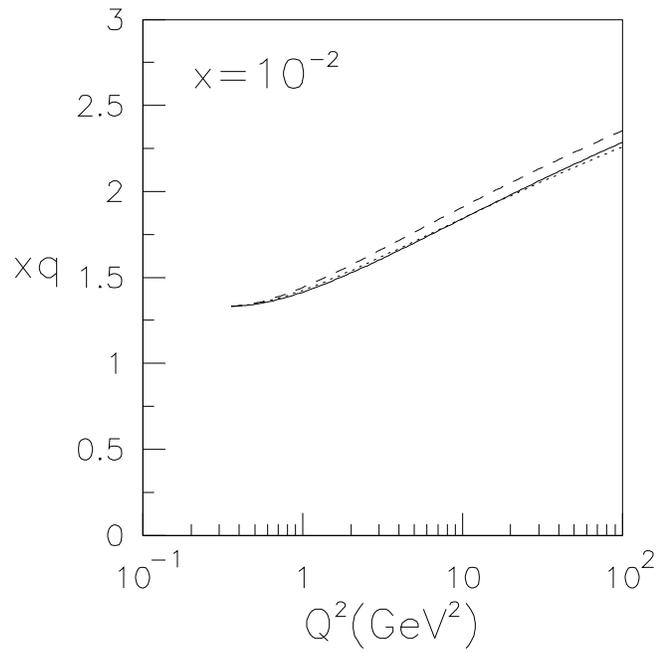
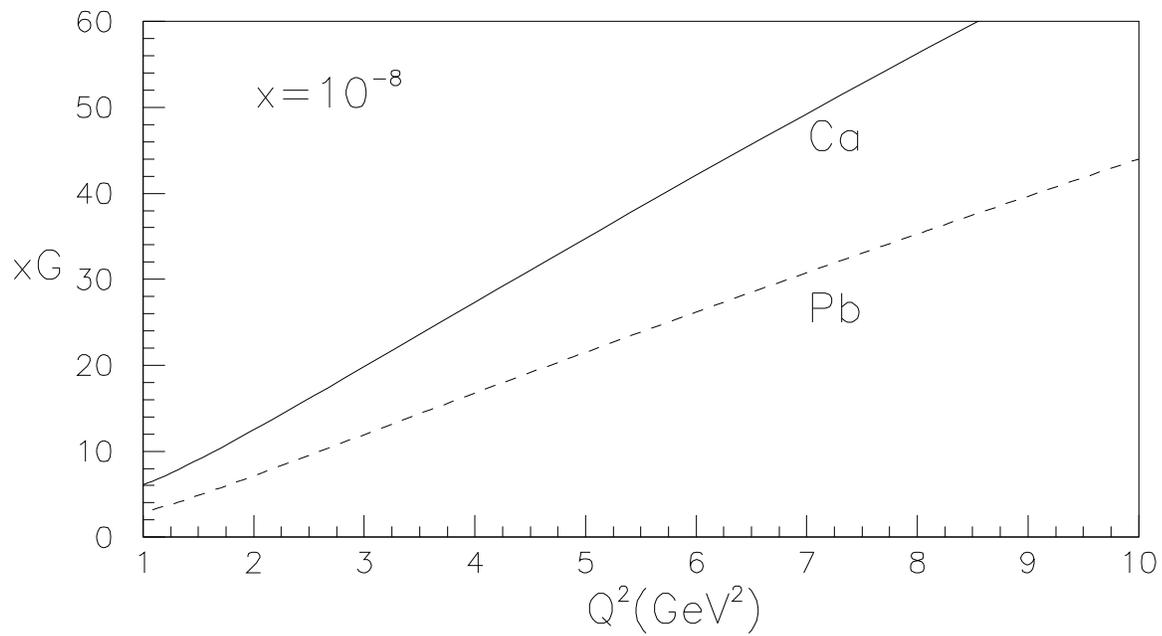
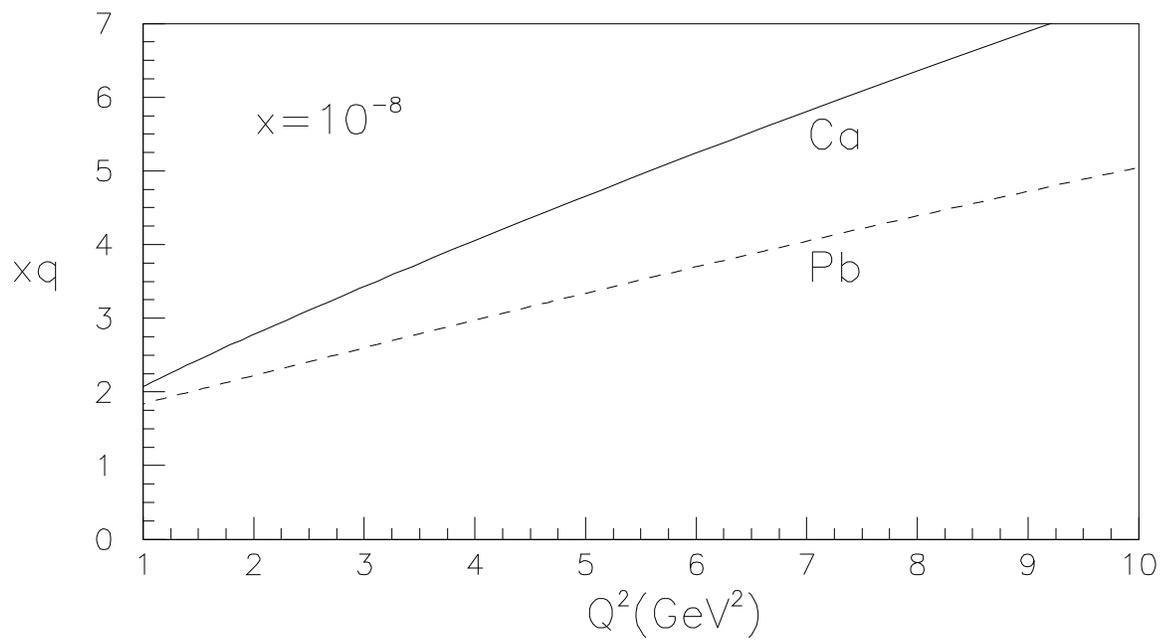
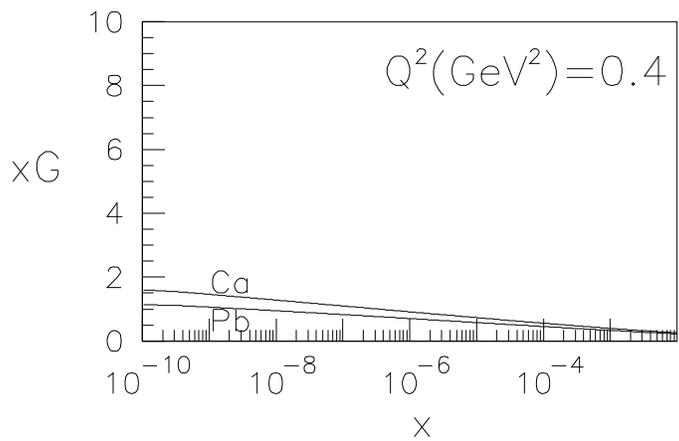
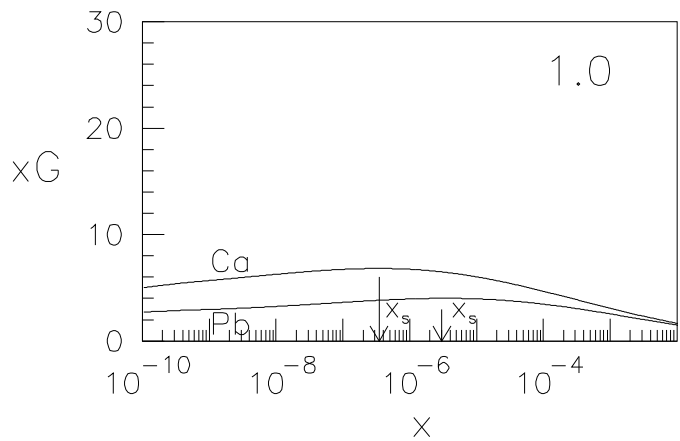
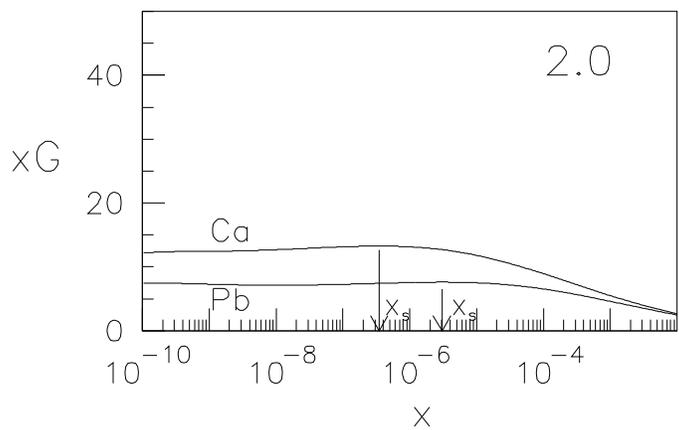
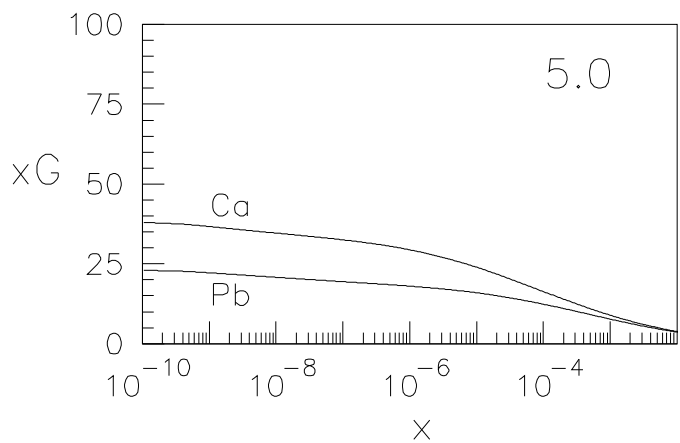
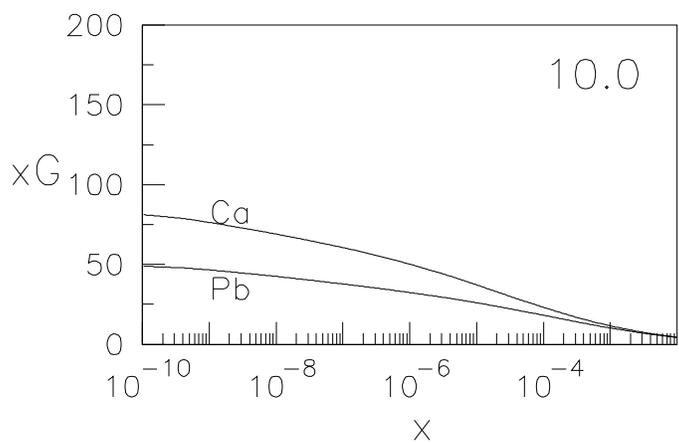
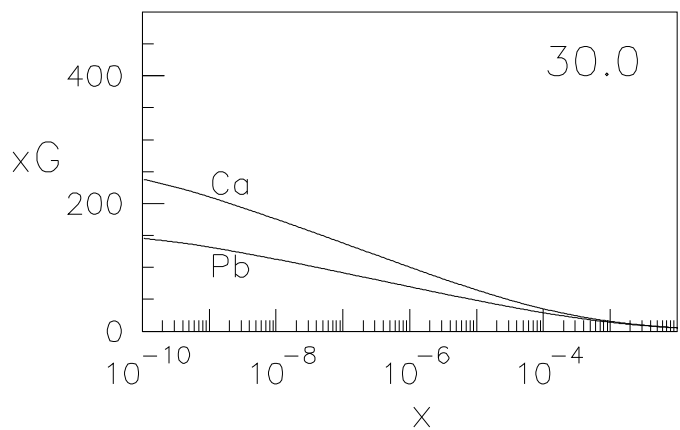
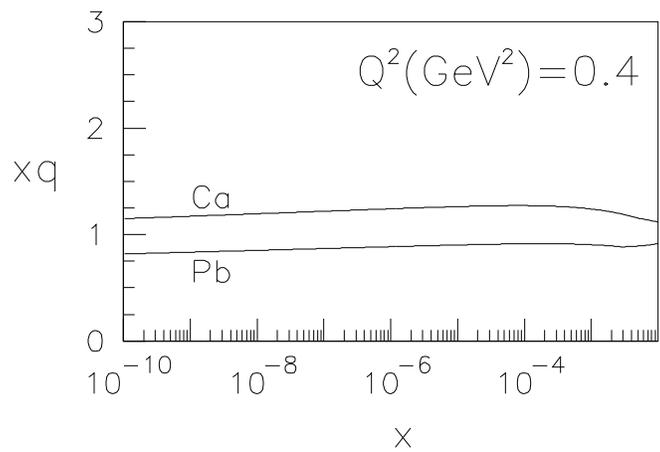
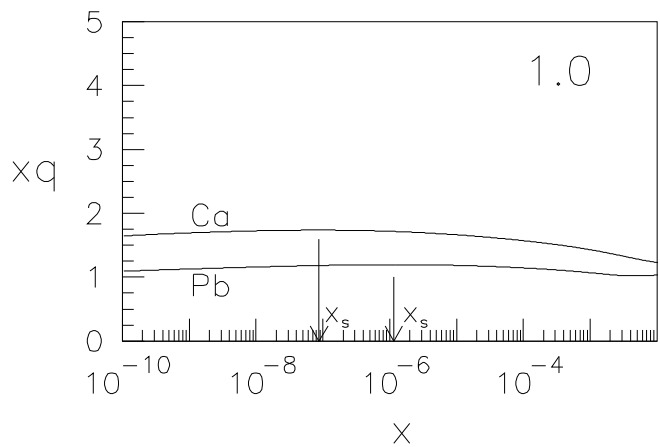
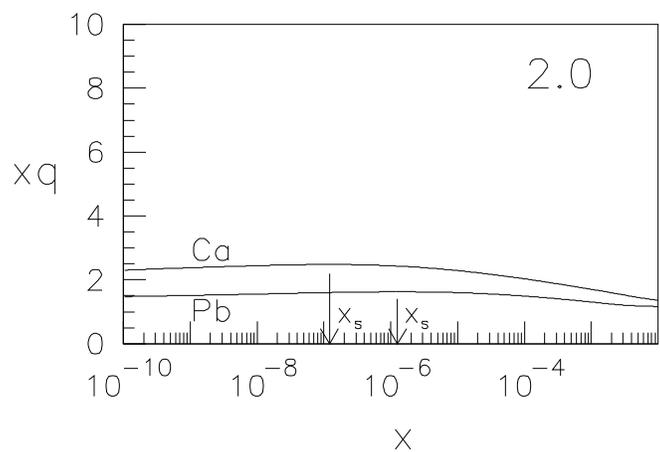
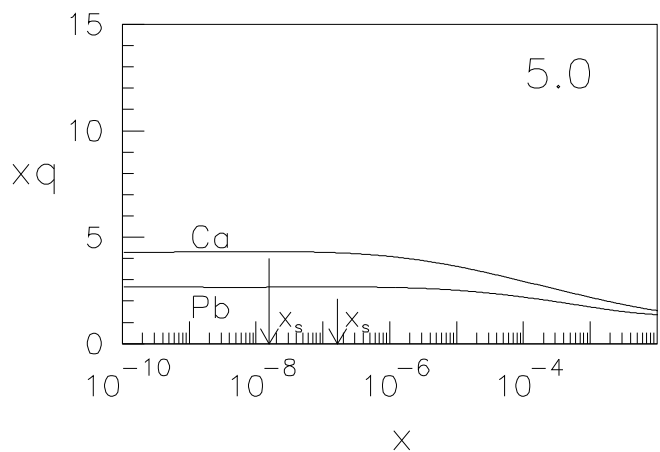
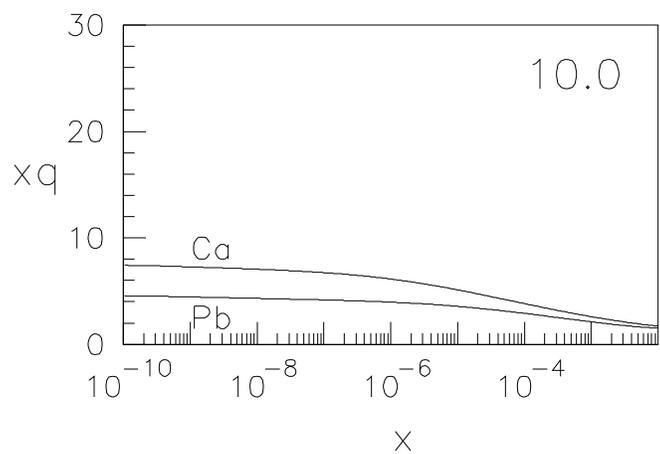
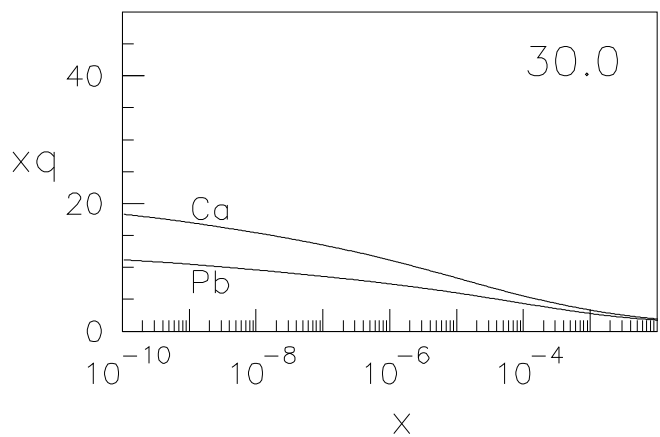
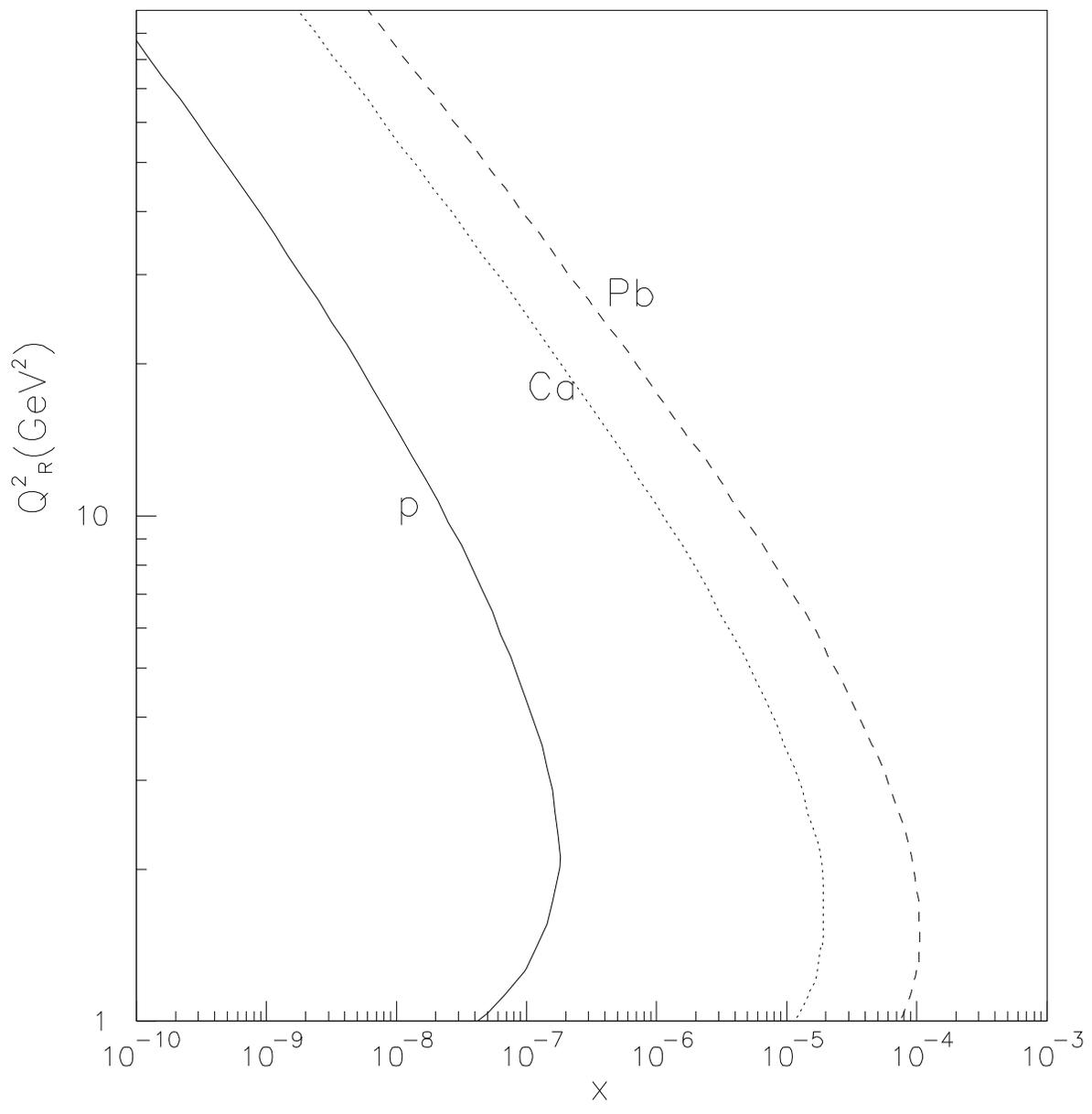
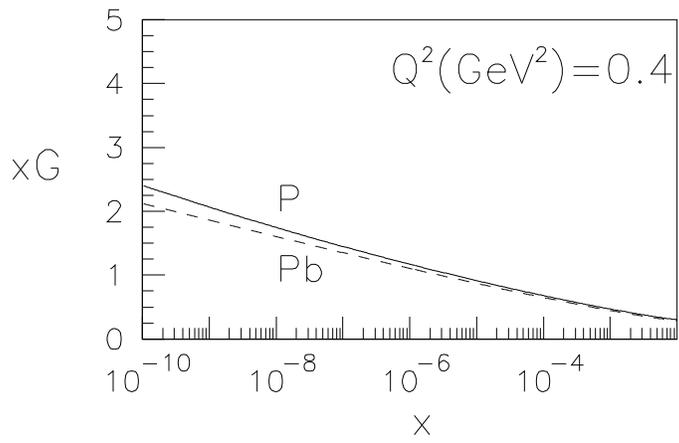
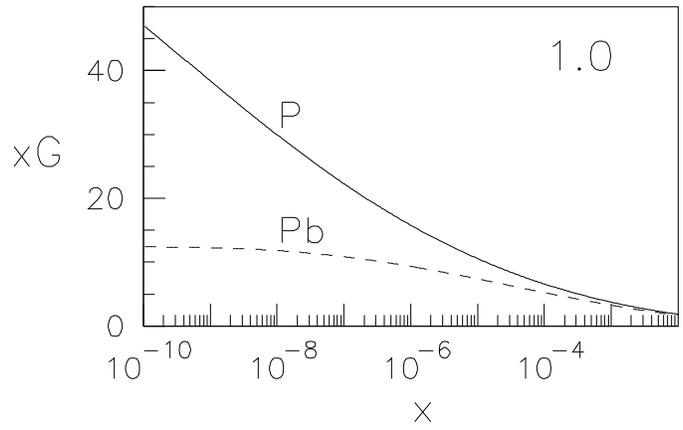
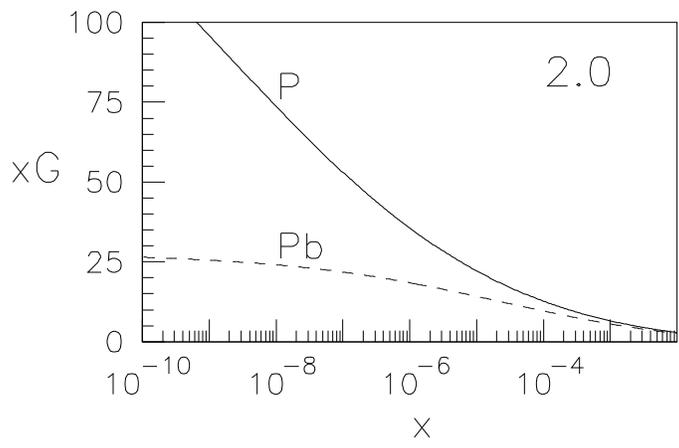
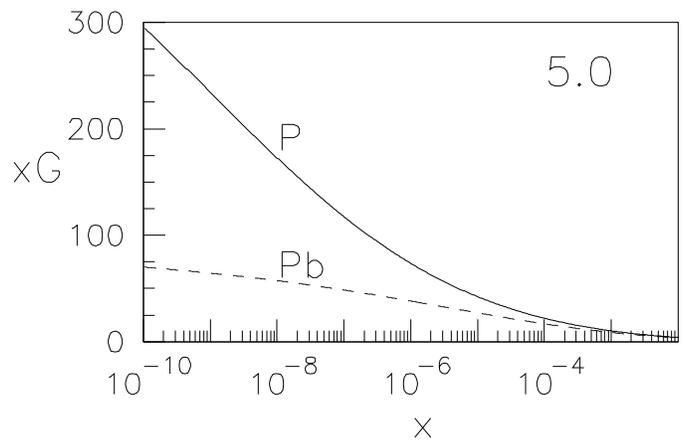
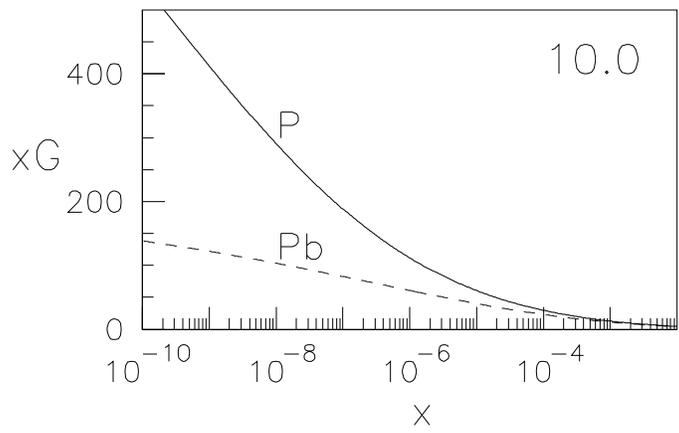
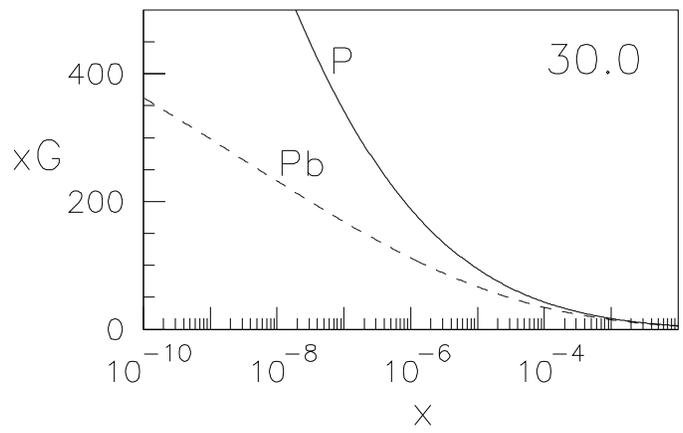
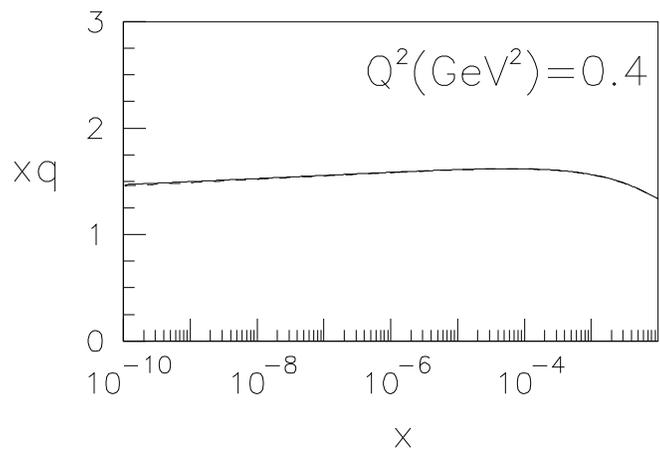
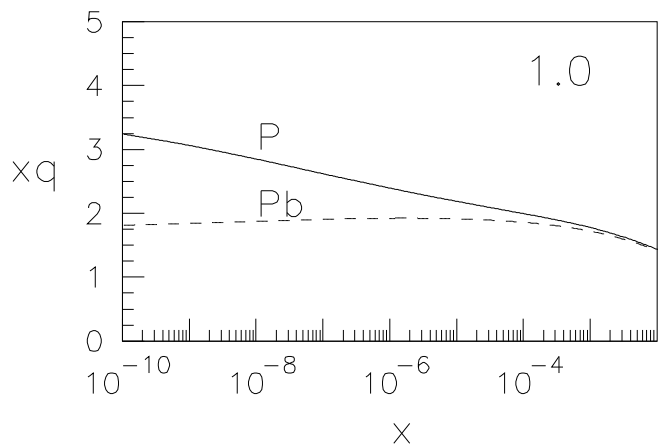
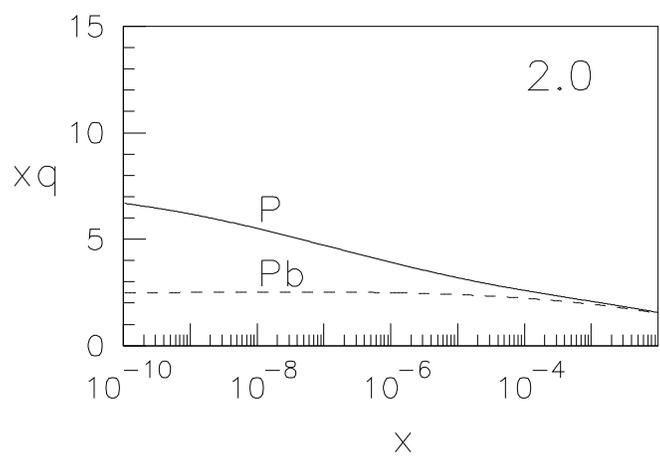
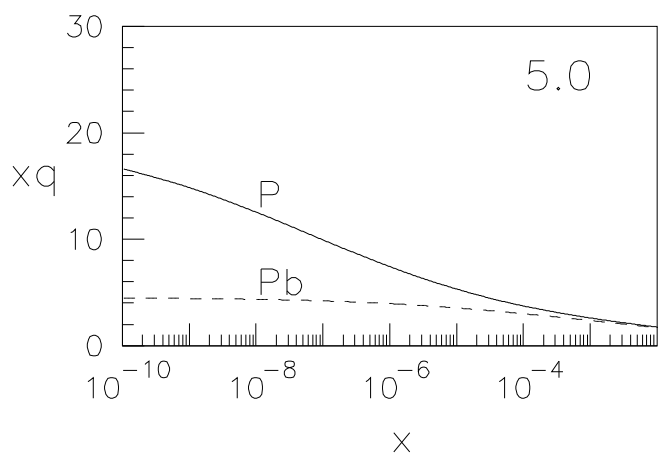
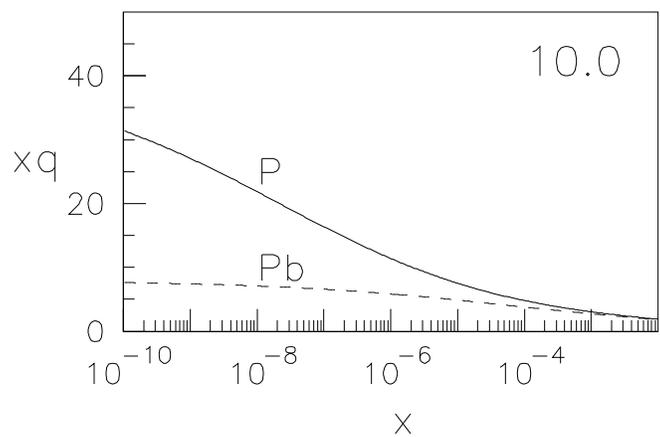
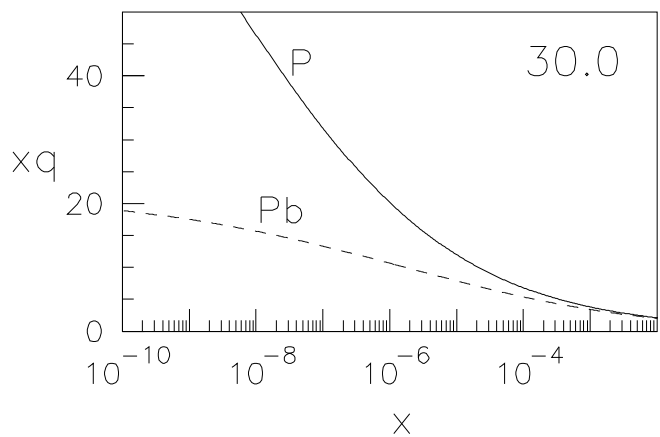
\end{document}